%% file: eett_main_P2.tex
\documentstyle[12pt,axodraw,epsfig]{article}
\input{eett_def}

\begin{document}
\newcommand{\chic}{\chi^*}
\def\href#1#2{#2}
\input{eett_title_P2}
\clearpage
\tableofcontents
\clearpage
\listoffigures
\listoftables    
\clearpage
\input{eett_text_P2}
\clearpage
\input{eett_refs_P2}
\end{document}

%% file: eett_def.tex
\newcommand{\eqnzero}{\setcounter{equation}{0}} 

\newcommand{\bq}{\begin{equation}}
\newcommand{\eq}{\end{equation}}
\newcommand{\ba}{\begin{eqnarray}}
\newcommand{\ea}{\end{eqnarray}}

\newcommand{\baa}[1]{\begin{array}{#1}}
\newcommand{\eaa}{\end{array}}

\newcommand{\nll}{\nonumber\\}

\newcommand{\Rz} {\mbox{$R_{\sss{Z}}  $}}

\newcommand{\Litwo}{\mbox{${\rm{Li}}_{2}$}}

\def\gz{\Gamma_{\sss{Z}}}

\def\Pgg{\Pi_{\gamma\gamma}}

\def\nl{\nonumber \\}

\def\href#1#2{#2}

\newcommand{\bqa}{\begin{eqnarray}}
\newcommand{\eqa}{\end{eqnarray}}
\newcommand{\ds }{\displaystyle}
\newcommand{\sss}[1]{\scriptscriptstyle{#1}}

\newcommand{\ph}{\gamma}
\newcommand{\ab}{A}
\newcommand{\zb}{Z}
\newcommand{\wb}{W}
\newcommand{\hb}{H}

\newcommand{\ff}{f}

\newcommand{\fep}{e^{+}}
\newcommand{\fem}{e^{-}}

\newcommand{\fu}{u}

\newcommand{\ft}{t}

\newcommand{\cff}[6]{C_{0}\big( #1,#2,#3;#4,#5,#6 \big) }    
\newcommand{\sman}{s}

\newcommand{\mwl }{M_{\sss{W}}}
\newcommand{\mzl }{M_{\sss{Z}}}

\newcommand{\mtl }{m_t}
\newcommand{\mbl }{m_b}
\newcommand{\mel }{m_e}

\newcommand{\wml} {M_{\sss{W}}}

\newcommand{\mzs}{M^2_{\sss{Z}}}

\newcommand{\mts}{m^2_{t}}

\newcommand{\mes}{m^2_e}

\newcommand{\mtq }{m^4_{t}}

\newcommand{\fbf}{{\bar{f}}}

\newcommand{\lpar}{\left(}                            
\newcommand{\rpar}{\right)}
\newcommand{\lrbr}{\left[}
\newcommand{\rrbr}{\right]}

\newcommand{\ib  }{i}
\newcommand{\qf  }{Q_f  }

\newcommand{\qe  }{Q_e  }

\newcommand{\qes }{Q^2_e}

\newcommand{\gdp }{\gamma_{+  }}
\newcommand{\gdm }{\gamma_{-  }}

\newcommand{\gdmu}{\gamma_{\mu}}

\newcommand{\vvertil}[3]{F^{#1}_{#2}\lpar{#3}\rpar}
\newcommand{\cvetril}[3]{{\cal{F}}^{#1}_{#2}\lpar{#3}\rpar}
\newcommand{\cvertil}[3]{{\cal{F}}^{#1}_{#2}\lpar{#3}\rpar}

\newcommand{\gadu}[1]{\gamma_{#1}}
\newcommand{\siw }{s_{\sss{W}}}           

\newcommand{\siws}{s^2_{\sss{W}}}
\newcommand{\cows}{c^2_{\sss{W}}}

\newcommand{\vpa}[2]{\sigma_{#1}^{#2}}
\newcommand{\vpae }{\sigma_e}
\newcommand{\vma}[2]{\delta_{#1}^{#2}}

\newcommand{\tcie}{I^{(3)}_e}

\newcommand{\tcit}{I^{(3)}_t}

\newcommand{\bff}[4]{B_{#1}\big( #2;#3,#4\big)}             
\newcommand{\fbff}[4]{B^{F}_{#1}\big(#2;#3,#4\big)}        
\newcommand{\scff}[1]{C_{#1}}             
\newcommand{\sdff}[1]{D_{#1}}

    \newcommand{\tHs}{\mu}

    \newcommand{\stws}{s^2_{\sss{W}}}
    \newcommand{\stwf}{s^4_{\sss{W}}}
    \newcommand{\ctws}{c^2_{\sss{W}}}
    \newcommand{\ctwf}{c^4_{\sss{W}}}

  \newcommand{\vmau }{\delta  _t}
  \newcommand{\vmae }{\delta  _e}

\newcommand{\fer}{\rm{fer}}

\newcommand{\eqn}[1]{Eq.~(\ref{#1})}
\newcommand{\eqns}[2]{Eqs.~(\ref{#1})--(\ref{#2})}

\newcommand{\eqnsc}[2]{Eqs.~(\ref{#1}) and (\ref{#2})}

\newcommand{\tbn}[1]{Table~\ref{#1}}

\newcommand{\tbnsc}[2]{Tabs.~\ref{#1} and \ref{#2}}
\newcommand{\fig}[1]{Fig.~\ref{#1}}

\newcommand{\qt}{Q_t}

\newcommand{\qu}{Q_t}

\newcommand{\qts}{Q^2_t}

\newcommand{\au}{a_t}
\newcommand{\vu}{v_t}

\newcommand{\Rxi}{R_{\gpar}}
\newcommand{\gpar}{\xi}

\newcommand{\sdfit} {\Delta_{4r}}
\newcommand{\sdtit} {\Delta_{3r}}

\newcommand{\tHlas}{\lambda^2}

\newcommand{\tmi }{t_-}
\newcommand{\tmis}{t^2_-}
\newcommand{\umi }{u_-}
\newcommand{\sqs}{\sqrt{s}}
\newcommand{\szmi}{s_-}
\newcommand{\szpl}{s_+}
\newcommand{\tpl}{t_+}

\newcommand{\jaat}{J_{\sss AA}(-s,-t;\mel,\mtl)}

\newcommand{\jazt}{J_{\sss ZA}(-s,-t;\mel,\mtl)}

\newcommand{\cesoeo}{\cff{-\mes}{-\mes}{-s}{0}{\mel}{0}}
\newcommand{\ceszeo}{\cff{-\mes}{-\mes}{-s}{\mzl}{\mel}{0}}
\newcommand{\ctszto}{\cff{-\mts}{-\mts}{-s}{\mzl}{\mtl}{0}}
\newcommand{\ctsoto}{\cff{-\mts}{-\mts}{-s}{0}{\mtl}{0}}
\newcommand{\cattze}{\cff{-\mts}{-\mes}{-t}{\mtl}{\mzl}{\mel}}

\newcommand{\cateot}{\cff{-\mes}{-\mts}{-t}{\mel}{0}{\mtl}}

\newcommand{\ctstot}{\cff{-\mts}{-\mts}{-s}{\mtl}{0}{\mtl}}
\newcommand{\boptot}{\bff{0p}{-\mts}{0}{\mtl}} 
\newcommand{\bopeoe}{\bff{0p}{-\mes}{0}{\mel}}
\newcommand{\bofeeo}{\fbff{0}{-\mes}{\mel}{0}}
\newcommand{\bofsee}{\fbff{0}{-s}{\mel}{\mel}}
\newcommand{\ceseoe}{\cff{-\mes}{-\mes}{-s}{\mel}{0}{\mel}}

\newcommand{\Faa}[1]{{\cal F}^{\sss AA}_{\sss #1}}
\newcommand{\Fza}[1]{{\cal F}^{\sss ZA}_{\sss #1}}

\newcommand{\bofsoo}{\fbff{0}{-s}{0}{0}}
\newcommand{\bofszo}{\fbff{0}{-s}{\mzl}{0}}
\newcommand{\boftzt}{\fbff{0}{-\mts}{\mzl}{\mtl}}
\newcommand{\boftet}{\fbff{0}{-t}{\mel}{\mtl}}

\newcommand{\bofstt}{\fbff{0}{-s}{\mtl}{\mtl}}
\newcommand{\boftto}{\fbff{0}{-\mts}{\mtl}{0}}
\newcommand{\bebeta}{\frac{(1+\beta^2)}{2 \beta}}      
\newcommand{\betatm}{\beta_-}
\newcommand{\betatp}{\beta_+}
\newcommand{\lelog}{{l}_e}
\newcommand{\tman}{t}
\newcommand{\uman}{u}

%% file: eett_title_P2.tex
\setcounter{page}{0}
\thispagestyle{empty}

\vspace*{-3cm}

\begin{flushright}
{\bf
JINR E2--2002--20 \\
 {\tt hep-ph/0202112}\\
  February 2002 
}
\end{flushright}
\vspace*{\fill}
\begin{center}

{\LARGE\bf
Further study of the $e^{+}e^{-}\to\ff \bar{f}$ process \\[2.5mm]
with the aid of {\tt CalcPHEP} system}
\vspace*{1.5cm}

{\bf 
A.~Andonov, D.~Bardin, S.~Bondarenko$^*$,  \\[1mm]
P.~Christova, L.~Kalinovskaya, and G.~Nanava
}

\vspace*{3mm}
{\normalsize
{\it Laboratory for Nuclear Problems, JINR,\\
$^*$ Bogoluobov Laboratory of Theoretical Physics, JINR,\\ 
\vspace*{2mm}
     ul. Joliot-Curie 6,
     RU-141980 Dubna, Russia}}
\vspace*{2cm}

\end{center}

\begin{abstract}
\noindent
In this paper we complete a description of calculation of the one-loop 
amplitude for $e^+e^-\to\ff\bar{f}$ process started in CERN-TH/2001-308.
This study is performed within the framework of the project {\tt CalcPHEP}.
Here we add QED subsets of the one-loop diagrams and the soft-photon contribution.
The formulae we derived are realized in two independent {\tt FORTRAN} codes,
{\tt eeffLib}, which was written in an old fashioned way, i.e. manually,
and another one, created automatically with an aid of {\tt s2n\_f} (symbols to numbers) 
software --- a part of  {\tt CalcPHEP} system.
We present a comprehensive comparison between the two our codes as well as with
the results existing in the world literature.
\end{abstract}

\vfill

\vspace*{1mm}
\bigskip
\footnoterule
\noindent
{\footnotesize \noindent
Work supported in part  by 
INTAS $N^{o}$ 00-00313.
\\
E-mails: andonov@nusun.jinr.ru, bardin@nusun.jinr.ru, bondarenko@jinr.ru \\ 
\phantom{XXXXX}
penchris@nusun.jinr.ru, kalinov@nusun.jinr.ru, nanava@nusun.jinr.ru
}

%% file: eett_text_P2.tex
\addcontentsline{toc}{section}{Introduction}
\section*{Introduction}
\label{Untroduction}
\eqnzero
  Recently, detail reports on process $e^+e^-\to t\bar{t} \to  6 \ff$  become an active subject 
for energy of future electron linear colliders.
This process will be one of main process and   
therefore must be theoretically studied profoundly (see for example the review ~\cite{Beneke:2000hk}).

  In this connection we consider a new  calculation
of $e^+e^-\to \ff \bar\ff $ process at the one-loop level
made with an aid of computer system {\tt CalcPHEP},
where all the calculations from the Lagrangians up to numbers
are going to be eventually automatized, (see~\cite{CalcPHEP:2000}).

 Electroweak (EW) parts have been calculated in~\cite{eett_subm}
and a very good agreement with FeynArts~\cite{Hahn:2000jm} and~\cite{Zeuthen:2001}
were found.
 
 In this paper we added lacking in~\cite{eett_subm} QED corrections. 
Our strategy in the descriptions of the QED part is the same as in our 
first paper; many definitions and notations from it are used here.
References to an equation of the first part will be denoted as {\bf (I.S.eq)} with {\bf S} and
{\bf eq} being Section and equation numbers of Ref.~\cite{eett_subm}, correspondingly. 

\vspace*{1mm}

This paper is organized in a similar fashion as~\cite{eett_subm}.
\vspace*{1mm}

In Section 1, we briefly remind the structure of one-loop amplitudes
\vspace*{1mm}

Section 2 contains explicit expressions for all the QED {\em building blocks}
which were not covered in~\cite{eett_subm}: QED vertices, AA and ZA boxes. 
\vspace*{1mm}

Section 3 contains the total scalar form factors of the one-loop amplitudes, now 
with all QED additions.
\vspace*{1mm}

In Section 4 we present explicit expressions for helicity amplitudes made of total 
scalar form factors at one-loop level.
\vspace*{1mm}

Section 5 is an Annex containing some additional expression for different QED contributions
that might be derived analytically. They are not in the main stream of our paper:
Lagrangian $\rightarrow$ scalar form factors $\rightarrow$ helicity amplitudes $\rightarrow$
one-loop differential cross-section. However, they are  
useful for pedagogical reasons, and their coding in complimentary {\tt FORTRAN} branches
of {\tt eeffLib} provided us with powerful internal cross-checks of our codes 
for numerical calculations.
Actually, {\tt eeffLib} version of February'2002 has three QED branches.
\vspace*{1mm}

Finally, Section 6 is a revised version of Section 5 of~\cite{eett_subm} in which
we present again results of a comprehensive numerical comparison between {\tt eeffLib} 
and {\tt ZFITTER}. The reason for this revision is due to debugging of the December'2001
version of {\tt eeffLib} resulting in a little change of our numbers beginning 4th or 5th digits.
In this paper we also present a comparison with our another code,  which was
created automatically using {\tt s2n\_f} software.
We also present a comprehensive comparison between the results derived with two our codes 
and the results existing in the world literature. 
In particular, we found a high precision agreement
with {\tt FeynArts} results up to 11 digits for the 
differential cross-sections with virtual corrections,
and with resent results of~\cite{Zeuthen:2001} within 7-8 digits even with soft photons included,
see~\cite{K-BZcomparison:2002}.

\section{Amplitudes}
\setcounter{equation}{0}

 We work in the $LQD$ basis, and the final-state fermion masses are not 
ignored as in previous~\cite{eett_subm}.
The electron mass is ignored everywhere, but arguments of logs. 
 Also we work in the $\Rxi$ gauge.
We checked the cancellation of $\xi$-dependent terms in three 
gauge-invariant subsets of diagrams separately.
The first subset is the so-called cluster in the QED sector (or $A$ cluster, see definitions 
below), the second and third are $AA$ boxes and $ZA$ boxes, correspondingly.  

 In the $LQD$ basis, the $\gamma$ and $\zb$ exchange one-loop amplitudes have the following 
structure:
\bqa
A^{\sss{\rm{IBA}}}_{\ph} = \ib\frac{4\pi\qe\qf}{s}
\alpha(\sman) \gadu{\mu} \otimes \gadu{\mu}\,,
\label{Born_modulo-old}
\eqa
and
\bqa
{\cal A}^{\sss{\rm{IBA}}}_{\sss{\zb}}&=&
\ib\,e^2
\frac{\chi_{\sss{Z}}(s)}{s}
   \Biggl\{
    \tcie\tcit
     \gadu{\mu} {\gdp } \otimes
       \gadu{\mu} {\gdp } 
    \vvertil{}{\sss{LL}}{s,t}      
+    \vmae\tcit
    \gadu{\mu}     
      \otimes \gadu{\mu} {\gdp} 
    \vvertil{}{\sss{QL}}{s,t} 
\nll &&
+\tcie\vmau \gadu{\mu}{\gdp}\otimes\gadu{\mu}  
    \vvertil{}{\sss{LQ}}{s,t}
+\vmae\vmau \gadu{\mu}\otimes\gadu{\mu}
    \vvertil{}{\sss{QQ}}{s,t}
\nll &&
+    \tcie\tcit
   \gadu{\mu}{\gdp}\otimes\lpar 
     -i m_t D_{\mu} \rpar     
    \vvertil{}{\sss{LD}}{s,t}
+\vmae\tcit \gadu{\mu} \otimes \lpar  
     -i m_t D_{\mu} \rpar  
    \vvertil{}{\sss{QD}}{s,t}
\Biggr\},\qquad
\label{structures-old}
\eqa
where {\it untilded} and {\it tilded} form factors are related by {\bf Eqs.~(I.1.11)}. Like
Part I, we present all the explicit expressions in term of {\it untilded} quantities.
Furthermore,
\bq
\alpha(s)=\frac{\alpha}
{\ds{1-\frac{\alpha}{4\pi}\Bigl[\Pgg^{\fer}(s)-\Pgg^{\fer}(0)\Bigr]}}
\label{alpha_fer-old}
\eq
is the fermionic component of the running QED coupling $\alpha(\sman)$ and 
\bqa
\chi_{\sss{Z}}(\sman)&=&\frac{1}{4\siws\cows}\frac{\sman}
{\ds{\sman - \mzs + \ib\frac{\gz}{\mzl}\sman}}\,
\label{propagators}
\eqa
is the $\zb/\ph$ propagator ratio with an $\sman$-dependent (or constant) $\zb$ width.
\section{Building Blocks (QED part)}
\eqnzero

\subsection{The $Zff$ and $\gamma ff$ vertices}
 First of all we have to add vertex QED building blocks to 
the scalar form factors of {\bf Eq.(I.2.60)} and
finally to the complete scalar form factors of {\bf Eqs.(I.3.117)}.

 The total vertex scalar form factors $\gamma\ft\bar{t}$ and $\zb\ft\bar{t}$ {\bf Eqs.~(I.2.60)}
are now sums over all bosonic contributions $B=\ab,\zb,\wb,\hb$,
since we add the diagram with virtual $\ph=\ab$.

 All the 24 components of the total form factors in the $LQD$ basis look like:
\bqa
  \vvertil{\gamma(\sss Z) tt}{\sss{L,Q,D}}{\sman}  =
  \vvertil{\gamma(\sss Z){\sss A}}{\sss{L,Q,D}}{\sman} 
+ \vvertil{\gamma(\sss Z){\sss Z}}{\sss{L,Q,D}}{\sman} 
+ \vvertil{\gamma(\sss Z){\sss W}}{\sss{L,Q,D}}{\sman}
+ \vvertil{\gamma(\sss Z){\sss H}}{\sss{L,Q,D}}{\sman},
\eqa
The $\ab$ cluster was formed using the same philosophy as in~\cite{eett_subm}, see
{\bf Eqs.~(I.2.54)-(I.2.59)}. Note, that $\vvertil{\gamma{\sss{A}}}{\sss{L}}{\sman}$ and
$\vvertil{\gamma{\sss{H}}}{\sss{L}}{\sman}$ are equal to zero.
\subsubsection{ Library of QED Form Factors for $Att$ clusters}
 
\begin{figure}[h]
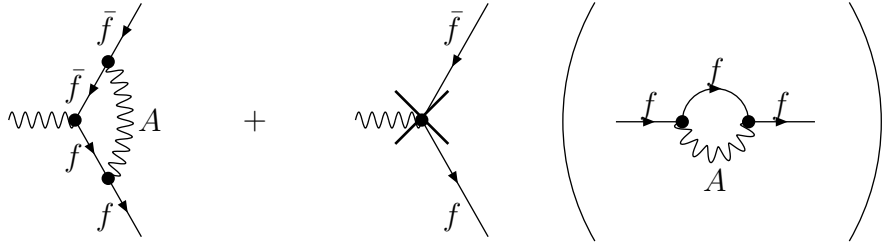

\caption
[$\ab$ cluster.] 
{$\ab$ cluster. 
One fermionic self-energy diagram in brackets
gives rise to the counter term contribution depicted by the solid cross.
\label{QEDcluster} }
\end{figure}

  Up to one-loop level, there are two diagrams, which contribute 
to the $\ab$ cluster, see ~\fig{QEDcluster}.
\vspace{-10mm}
\[
\baa{ccccc}
\begin{picture}(55,88)(0,41)
  \Photon(0,44)(25,44){3}{5}
  \Vertex(25,44){2.5}
  \PhotonArc(0,44)(44,-28,28){3}{9}
  \Vertex(37.5,22){2.5}
  \Vertex(37.5,66){2.5}
  \ArrowLine(50,88)(37.5,66)
  \ArrowLine(37.5,66)(25,44)
  \ArrowLine(25,44)(37.5,22)
  \ArrowLine(37.5,22)(50,0)
  \Text(33,75)[lb]{$\fbf$}
  \Text(21,53)[lb]{$\fbf$}
  \Text(49,44)[lc]{$A$}
  \Text(21,35)[lt]{$\ff$}
  \Text(33,13)[lt]{$\ff$}
  \Text(1,1)[lb]{}
\end{picture}
\qquad  &+& \qquad
\begin{picture}(55,88)(0,41)
  \Photon(0,44)(25,44){3}{5}
  \Vertex(25,44){2.5}
  \ArrowLine(50,88)(25,44)
  \ArrowLine(25,44)(50,0)
  \Text(33,75)[lb]{$\fbf$}
  \Text(33,13)[lt]{$\ff$}
  \Text(1,1)[lb]{}
\SetScale{2.0}
  \Line(17.5,17.5)(7.5,27.5)
  \Line(7.5,17.5)(17.5,27.5)
\end{picture}
&& \qquad 
\begin{picture}(75,20)(0,7.5)
\Text(12.5,13)[bc]{$\ff$}
\Text(62.5,13)[bc]{$\ff$}
\Text(37.5,26)[bc]{$\ff$}
\Text(37.5,-8)[tc]{$A$}
\Text(0,-7)[lt]{$ $}
  \ArrowLine(0,10)(25,10)
  \CArc(10,10)(90, -30, 30)
  \CArc(70,10)(90, -210 ,-150)
  \ArrowArcn(37.5,10)(12.5,180,0)
  \PhotonArc(37.5,10)(12.5,180,0){3}{7}
  \Vertex(25,10){2.5}
  \Vertex(50,10){2.5}
  \ArrowLine(50,10)(75,10)
\end{picture}
\eaa
\]

\vspace{10mm}
Since after wave function renormalization, the scalar form factors became UV-finite,
instead of {\bf Eq.~(I.2.61)}, we have for all 6 form factors which are also separately
gauge-invariant:
\bqa
F^{\gamma(z){\sss A}}_{\sss I} = {\cal F}^{\gamma(z){\sss A}}_{\sss I},
\eqa
where $I = L,Q,D$. Individual components are:
\bqa
{\cal F}^{\gamma{\sss A}}_{\sss L} &=& 0,
\nll
{\cal F}^{\gamma{\sss A}}_{\sss Q} &=& \qu^2 \stws \biggl\{
  2 \lpar s-2 \mts\rpar  \ctstot
\nll &&
 -3 \bofstt + 3 \boftto  - 4 \mts \boptot \biggr\},
\nll
{\cal F}^{\gamma{\sss A}}_{\sss D} &=&
-\frac{\qu^3\stws}{\tcit} \frac{4}{\sdtit}\bigg[ \bofstt-\boftto \bigg],
\nll 
{\cal F}^{z{\sss A}}_{\sss L} &=& {\cal F}^{\gamma \sss A}_{Q} + \qu^2 \stws 
    \frac{8\mts}{\sdtit}\bigg[ \bofstt-\boftto\bigg],
\nll
{\cal F}^{z{\sss A}}_{\sss Q} &=& 
{\cal F}^{\gamma \sss A}_{Q} - \qu^2 \stws 
    \frac{8\mts}{\sdtit} \frac{\tcit}{\delta_t} \bigg[ \bofstt-\boftto \bigg],
\nll
{\cal F}^{z{\sss A}}_{\sss D} &=& 
-\frac{\qu^2\stws}{\tcit}\frac{2\vu}{\sdtit} \bigg[ \bofstt-\boftto \bigg],
\eqa
with
\bqa
\sdtit&=& 4 \mts - \sman\,.
\eqa

\subsubsection{Scalar form factor for electron case}
$Aee$ cluster is described by only one scalar form factor: 
\bqa
{\cal F}^{{\sss A},e} \lpar s \rpar &=& Q_e^2 \stws \bigg[ 2 s \ceseoe
\\ &&
  - 3 \bofsee + 3 \bofeeo - 4 \mes \bopeoe \bigg].
\nonumber
\eqa
\subsection{Amplitudes of QED boxes}
The contributions of QED $AA$ and $ZA$ boxes form gauge-invariant and UV finite subsets.
In terms of six structures  $(L,R) \otimes (L,R,D)$ they read:
\bqa
\bigg({\cal B}^{\sss AA(ZA)} \bigg)^{d+c}\hspace*{-5.5mm} & = &
\label{anyboxamplitude}
k^{\sss AA(ZA)}_{\rm norm}\frac{g^4}{s}
\Biggl[
    \lrbr  \gdmu \gdp \otimes  \gdmu \gdp \rrbr
{\cal F}^{\sss AA(ZA)}_{\sss LL} \lpar s,t,u \rpar
  + \lrbr  \gdmu \gdp \otimes  \gdmu \gdm  \rrbr
{\cal F}^{\sss AA(ZA)}_{\sss LR}\lpar s,t,u \rpar 
\nll[1mm] 
  &+& \lrbr  \gdmu \gdm \otimes  \gdmu \gdp \rrbr 
{\cal F}^{\sss AA(ZA)}_{\sss RL}\lpar s,t,u \rpar
  + \lrbr  \gdmu \gdm   \otimes  \gdmu \gdm \rrbr
{\cal F}^{\sss AA(ZA)}_{\sss RR} \lpar s,t,u \rpar
\\[1mm] 
  &+& \lrbr \gdmu \gdp  \otimes \lpar -i \mtl I D_\mu \rpar \rrbr
{\cal F}^{\sss AA(ZA)}_{\sss LD}\lpar s,t,u \rpar
  + \lrbr \gdmu \gdm  \otimes \lpar -i \mtl I D_\mu \rpar \rrbr 
{\cal F}^{\sss AA(ZA)}_{\sss RD}\lpar s,t,u \rpar
\Biggr].
\nonumber
\eqa
where for shortening of presentation we factorize out normalization factors: 
\bq
k^{\sss AA}_{\rm norm}= \stwf Q_e^2 Q_t^2\,,
\qquad
k^{\sss ZA}_{\rm norm}= \frac{\stws Q_e Q_t}{\ctws}\,.
\eq
For completeness and subsequent use we remind $k^{\sss ZZ}_{\rm norm}$ appearing in 
{\bf Eq.~(I.2.95)}:
\bq
k^{\sss ZZ}_{\rm norm}= \frac{1}{32\ctwf}\,.
\eq
\subsubsection{ $AA$-box contribution}
 There are only two $AA$ diagrams, {\em direct} and {\em crossed}:

\begin{figure}[!h]
\vspace*{-23mm}
\[
\hspace*{-2cm}
{
\begin{array}{ccc}
\begin{picture}(100,105)(0,49.5)
  \ArrowLine(0,15)(37.5,15)
  \ArrowLine(37.5,15)(37.5,90)
  \ArrowLine(37.5,90)(0,90)
  \Photon(37.5,15)(112.5,15){4}{15}
  \Photon(37.5,90)(112.5,90){4}{15}
  \Vertex(37.5,15){2.5}
  \Vertex(37.5,90){2.5}
  \Vertex(112.5,15){2.5}
  \Vertex(112.5,90){2.5}
  \ArrowLine(150,90)(112.5,90)
  \ArrowLine(112.5,90)(112.5,15)
  \ArrowLine(112.5,15)(150,15)
\Text(18.75,105)[tc]{$e^+$}
  \Text(75,108.75)[tc]{$A$}
  \Text(131.25,110.5)[tc]{$\bar t$}
  \Text(18.75,52.5)[lc]{$e$}
  \Text(131.8,52.5)[lc]{$t$}
  \Text(18.75,0)[cb]{$e^-$}
  \Text(75,-3.75)[bc]{$A$}
  \Text(131.25,-3)[cb]{$t$}
\end{picture}
\qquad \qquad \qquad 
&+&
\quad 
\begin{picture}(100,105)(0,49.5)
  \ArrowLine(0,15)(37.5,15)
  \ArrowLine(37.5,15)(37.5,90)
  \ArrowLine(37.5,90)(0,90)
  \Photon(37.5,15)(112.5,90){4}{20}
  \Photon(37.5,90)(112.5,15){4}{20}
  \Vertex(37.5,15){2.5}
  \Vertex(37.5,90){2.5}
  \Vertex(112.5,15){2.5}
  \Vertex(112.5,90){2.5}
  \ArrowLine(150,90)(112.5,90)
  \ArrowLine(112.5,90)(112.5,15)
  \ArrowLine(112.5,15)(150,15)
\Text(18.75,105)[tc]{$e^+$}
  \Text(75,108.75)[tc]{$A$}
  \Text(131.25,110.5)[tc]{$\bar t$}
  \Text(18.75,52.5)[lc]{$e$}
  \Text(131.8,52.5)[lc]{$t$}
  \Text(18.75,0)[cb]{$e^-$}
  \Text(75,-3.75)[bc]{$A$}
  \Text(131.25,-3)[cb]{$t$}
\end{picture}
\end{array}
}
\]
\vspace{10mm}
\caption{Direct and crossed $AA$ boxes. \label{AA_box_dc} }
\end{figure}
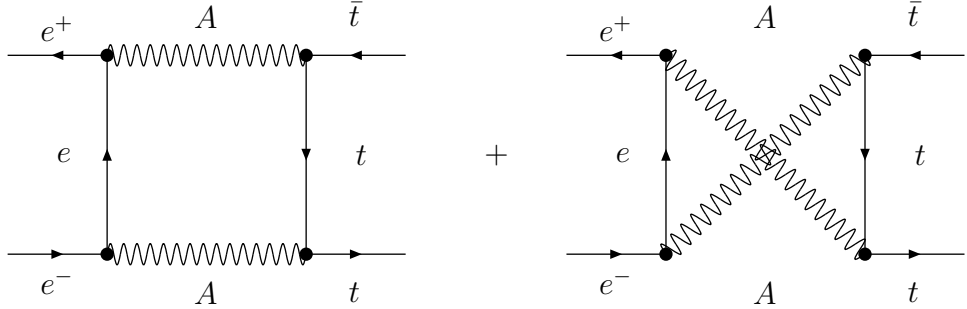

The six form factors of $AA$ boxes might be expressed in terms of only
four auxiliary functions ${\cal F}_1$ and ${\cal H}_{1,2,3}$:
\bqa
 \Faa{LL}\lpar s,t,u \rpar &=&  \Faa{RR}\lpar s,t,u \rpar
                   = {\cal H}_1 \lpar s,t \rpar - {\cal H}_1 \lpar s,u \rpar 
            + {\cal H}_2 \lpar s,t \rpar + {\cal H}_3 \lpar s,u \rpar,
\nll 
 \Faa{LR}\lpar s,t,u \rpar &=& \Faa{RL}\lpar s,t,u \rpar
                   = {\cal H}_1 \lpar s,t \rpar - {\cal H}_1 \lpar s,u \rpar 
            - {\cal H}_2 \lpar s,u \rpar - {\cal H}_3 \lpar s,t \rpar,
\nll
 \Faa{LD}\lpar s,t,u \rpar &=& \Faa{LD}\lpar s,t,u \rpar
                   = {\cal F}_1 \lpar s,t \rpar - {\cal F}_1\lpar s,u \rpar.
\eqa
The auxiliary functions are rather short:
\bqa
{\cal F}_1 \lpar s,t \rpar &=& -\frac{1}{2} \frac{s}{\sdfit} 
                \Biggl\{\frac{1}{\sdfit} \Bigg( - \tmi^3  \jaat           
\nll &&
               +t s \,\bigg[ s \, \cesoeo
               +\lpar s - 2  \mts \rpar \ctsoto \bigg] \Bigg)
\nll &&
       + 2  \frac{t}{\sdtit} \bigg[2  \mts \ctsoto + \boftto - \bofsoo \bigg]
\nll &&
       + 2  \frac{t}{\tmi} \bigg[  \boftet - \boftto \bigg]  \Biggr\}\,,
\\[1mm]
{\cal H}_1 \lpar s,t \rpar &=&  - \tmi \bigg[ \frac{1}{2} \jaat-\cateot\bigg]
\nll &&
      + \frac{s}{4 \sdfit} \Biggl\{ \tmi \lpar t+\frac{\tpl\tmi^2}{\sdfit} \rpar \jaat
\nll &&
      + 2  \mts \lpar 1 -2  \frac{t}{\tmi}\rpar \bigg[ \boftet-\boftto \bigg] \Biggr\}\,,
\\[1mm]
{\cal H}_2 \lpar s,t \rpar &=& 
\frac{ s}{4\sdfit}  \Biggl\{
    \bigg[ - 2  \mts s + \lpar s - 4 \mts \rpar
    \bigg(s + 2\tmi- \lpar  s\tpl + 2 t\tmi \rpar \frac{s}{\sdfit}\bigg) \bigg] 
\nll &&
\times \ctsoto
\nll &&
        + \lpar s + 2\tmi \rpar \lpar 1 - \tpl \frac{s}{\sdfit}\rpar  s \, \cesoeo
\nll && 
       - 2  \tmi  \bigg[ \boftet - \bofsoo \bigg] 
\nll &&
       - 4  \mts  \bigg[ \boftet - \boftto \bigg]  \Biggr\} \, ,
\\[1mm]
{\cal H}_3 \lpar s,t \rpar &=& \frac{s}{4 \sdfit^2} \lpar s + 2 \tmi \rpar \tmi^3 \jaat\,.
\eqa
Here
\bq
\sdfit=-\tman\uman + \mtq\,,
\eq
and $J^{\sss AA}\lpar Q^2,P^2;M_1,M_2 \rpar$ is due to a procedure of disentengling
of the infrared divergences from $\sdff{0}$.
Its explicit expression reads $(P^2 > 0, Q^2 < 0$, and $M_1$ is ignored everywhere but $\ln$):
\bqa 
J^{\sss AA}\lpar Q^2,P^2;M_1,M_2 \rpar &\hspace*{-2.3mm} =&\hspace*{-2.3mm} 
\frac{1}{ P_2+M_2^2}
         \Biggl\{
       \ln\frac{\lpar P^2+M^2_2 \rpar^2}{-Q^2 P^2}\ln\lpar \frac{P^2}{-Q^2}\rpar
      -\frac{1}{2}\ln^2\lpar\frac{M_1^2}{-Q^2}\rpar
      -\frac{1}{2}\ln^2\lpar\frac{M_2^2}{-Q^2}\rpar
\nll &&\hspace*{-1mm}
      + \ln^2\lpar 1+\frac{M^2_2}{P^2} \rpar
           -2 \Litwo \lpar\frac{P^2}{P^2+M^2_2} \rpar 
+i \pi
           \ln\lrbr \frac{ \lpar P^2+M_2^2 \rpar^2}{M_1^2 M_2^2}\rrbr
        \Biggr\}.
\label{JAA}
\eqa 
Moreover, the relevant infrared divergent $\scff{0}$ function ($P^2 > 0$ again), is
\bqa 
C_0^{\sss {\rm IR}}
\lpar -M_1^2, -M_2^2, P^2; M_1, \lambda, M_2 \rpar
&\hspace*{-2mm} =&\hspace*{-2mm}
\frac{1}{2\lpar P^2+M^2_2 \rpar}
         \Bigg\{
       \ln\lrbr\frac{\lpar P^2+M_2^2\rpar^2}{ M_1^2 M_2^2}\rrbr
             \ln \frac{ P^2}{\tHlas} 
      - 2\Litwo\lpar\frac{P^2}{P^2+M_2^2}\rpar
\nll &&
      -\frac{1}{2} \ln^2\lpar \frac{M_1^2}{P^2}\rpar
      -\frac{1}{2} \ln^2\lpar \frac{M_2^2}{P^2}\rpar 
     + \ln^2\lpar 1+\frac{M_2^2}{P^2}\rpar
           \Bigg\}.
\eqa
\subsubsection{$ZA$ box contribution}
In $\Rxi$ gauge there are eight $ZA$ boxes, however, since electron mass is ignored,
only four diagrams without $\phi_0$ contribute: 

\begin{figure}[!h]
\vspace{-22mm}
\[
\hspace*{-2cm}
{
\begin{array}{ccc}
\begin{picture}(100,105)(0,49.5)
  \ArrowLine(0,15)(37.5,15)
  \ArrowLine(37.5,15)(37.5,90)
  \ArrowLine(37.5,90)(0,90)
  \Photon(37.5,15)(112.5,15){4}{15}
  \Photon(37.5,90)(112.5,90){4}{10}
  \Vertex(37.5,15){2.5}
  \Vertex(37.5,90){2.5}
  \Vertex(112.5,15){2.5}
  \Vertex(112.5,90){2.5}
  \ArrowLine(150,90)(112.5,90)
  \ArrowLine(112.5,90)(112.5,15)
  \ArrowLine(112.5,15)(150,15)
\Text(18.75,105)[tc]{$e^+$}
  \Text(75,108.75)[tc]{$Z$}
  \Text(131.25,112.5)[tc]{$\bar t$}
  \Text(18.75,52.5)[lc]{$e$}
  \Text(131.8,52.5)[lc]{$t$}
  \Text(18.75,0)[cb]{$e^-$}
  \Text(75,-3.75)[bc]{$A$}
  \Text(131.25,-3)[cb]{$t$}
\end{picture}
\qquad \qquad \qquad 
&+&
\quad 
\begin{picture}(100,105)(0,49.5)
  \ArrowLine(0,15)(37.5,15)
  \ArrowLine(37.5,15)(37.5,90)
  \ArrowLine(37.5,90)(0,90)
  \Photon(37.5,15)(112.5,15){4}{10}
  \Photon(37.5,90)(112.5,90){4}{15}
  \Vertex(37.5,15){2.5}
  \Vertex(37.5,90){2.5}
  \Vertex(112.5,15){2.5}
  \Vertex(112.5,90){2.5}
  \ArrowLine(150,90)(112.5,90)
  \ArrowLine(112.5,90)(112.5,15)
  \ArrowLine(112.5,15)(150,15)
\Text(18.75,105)[tc]{$e^+$}
  \Text(75,108.75)[tc]{$A$}
  \Text(131.25,112.5)[tc]{$\bar t$}
  \Text(18.75,52.5)[lc]{$e$}
  \Text(131.8,52.5)[lc]{$t$}
  \Text(18.75,0)[cb]{$e^-$}
  \Text(75,-3.75)[bc]{$Z$}
  \Text(131.25,-3)[cb]{$t$}
\end{picture}
\\ \\ 
\begin{picture}(100,105)(0,49.5)
  \ArrowLine(0,15)(37.5,15)
  \ArrowLine(37.5,15)(37.5,90)
  \ArrowLine(37.5,90)(0,90)
  \Photon(37.5,15)(112.5,15){4}{15}
\Line(37.5,90)(42.5,90)
\Line(47.5,90)(52.5,90)
\Line(57.5,90)(62.5,90)
\Line(67.5,90)(72.5,90)
\Line(77.5,90)(82.5,90)
\Line(87.5,90)(92.5,90)
\Line(97.5,90)(102.5,90)
\Line(107.5,90)(112.5,90)
  \Vertex(37.5,15){2.5}
  \Vertex(37.5,90){2.5}
  \Vertex(112.5,15){2.5}
  \Vertex(112.5,90){2.5}
  \ArrowLine(150,90)(112.5,90)
  \ArrowLine(112.5,90)(112.5,15)
  \ArrowLine(112.5,15)(150,15)
\Text(18.75,105)[tc]{$e^+$}
  \Text(75,108.75)[tc]{$\phi^0$}
  \Text(131.25,112.5)[tc]{$\bar t$}
  \Text(18.75,52.5)[lc]{$e$}
  \Text(131.8,52.5)[lc]{$t$}
  \Text(18.75,0)[cb]{$e^-$}
  \Text(75,-3.75)[bc]{$A$}
  \Text(131.25,-3)[cb]{$t$}
\end{picture}
\qquad \qquad \qquad
&+&
\quad 
\begin{picture}(100,105)(0,49.5)
  \ArrowLine(0,15)(37.5,15)
  \ArrowLine(37.5,15)(37.5,90)
  \ArrowLine(37.5,90)(0,90)
\Line(37.5,15)(42.5,15)
\Line(47.5,15)(52.5,15)
\Line(57.5,15)(62.5,15)
\Line(67.5,15)(72.5,15)
\Line(77.5,15)(82.5,15)
\Line(87.5,15)(92.5,15)
\Line(97.5,15)(102.5,15)
\Line(107.5,15)(112.5,15)
  \Photon(37.5,90)(112.5,90){4}{15}
  \Vertex(37.5,15){2.5}
  \Vertex(37.5,90){2.5}
  \Vertex(112.5,15){2.5}
  \Vertex(112.5,90){2.5}
  \ArrowLine(150,90)(112.5,90)
  \ArrowLine(112.5,90)(112.5,15)
  \ArrowLine(112.5,15)(150,15)
  \Text(18.75,105)[tc]{$e^+$}
  \Text(75,108.75)[tc]{$A$}
  \Text(131.25,112.5)[tc]{$\bar t$}
  \Text(18.75,52.5)[lc]{$e$}
  \Text(131.8,52.5)[lc]{$t$}
  \Text(18.75,0)[cb]{$e^-$}
  \Text(75,-3.75)[bc]{$\phi^0$}
  \Text(131.25,-3)[cb]{$t$}
\end{picture}
\\  \\  
\begin{picture}(100,105)(0,49.5)
  \ArrowLine(0,15)(37.5,15)
  \ArrowLine(37.5,15)(37.5,90)
  \ArrowLine(37.5,90)(0,90)
  \Photon(37.5,15)(112.5,90){4}{20}
  \Photon(37.5,90)(112.5,15){4}{10}
  \Vertex(37.5,15){2.5}
  \Vertex(37.5,90){2.5}
  \Vertex(112.5,15){2.5}
  \Vertex(112.5,90){2.5}
  \ArrowLine(150,90)(112.5,90)
  \ArrowLine(112.5,90)(112.5,15)
  \ArrowLine(112.5,15)(150,15)
\Text(18.75,105)[tc]{$e^+$}
  \Text(75,108.75)[tc]{$Z$}
  \Text(131.25,112.5)[tc]{$\bar t$}
  \Text(18.75,52.5)[lc]{$e$}
  \Text(131.8,52.5)[lc]{$t$}
  \Text(18.75,0)[cb]{$e^-$}
  \Text(75,-3.75)[bc]{$A$}
  \Text(131.25,-3)[cb]{$t$}
\end{picture}
\qquad \qquad \qquad
&+&
\quad 
\begin{picture}(100,105)(0,49.5)
  \ArrowLine(0,15)(37.5,15)
  \ArrowLine(37.5,15)(37.5,90)
  \ArrowLine(37.5,90)(0,90)
  \Photon(37.5,15)(112.5,90){4}{10}
  \Photon(37.5,90)(112.5,15){4}{20}
  \Vertex(37.5,15){2.5}
  \Vertex(37.5,90){2.5}
  \Vertex(112.5,15){2.5}
  \Vertex(112.5,90){2.5}
  \ArrowLine(150,90)(112.5,90)
  \ArrowLine(112.5,90)(112.5,15)
  \ArrowLine(112.5,15)(150,15)
\Text(18.75,105)[tc]{$e^+$}
  \Text(75,108.75)[tc]{$A$}
  \Text(131.25,112.5)[tc]{$\bar t$}
  \Text(18.75,52.5)[lc]{$e$}
  \Text(131.8,52.5)[lc]{$t$}
  \Text(18.75,0)[cb]{$e^-$}
  \Text(75,-3.75)[bc]{$Z$}
  \Text(131.25,-3)[cb]{$t$}
\end{picture}
\\  \\  
\begin{picture}(100,105)(0,49.5)
  \ArrowLine(0,15)(37.5,15)
  \ArrowLine(37.5,15)(37.5,90)
  \ArrowLine(37.5,90)(0,90)
  \Photon(37.5,15)(112.5,90){4}{20}
 \Line(37.5,90)(41.5,88)
 \Line(45.5,84)(49.5,80)
 \Line(53.5,76)(57.5,72)
 \Line(61.5,68)(65.5,64)
 \Line(69.5,60)(73.5,56)
 \Line(77.5,52)(81.5,48)
 \Line(85.5,44)(89.5,40)
 \Line(93.5,36)(97.5,32)
 \Line(101.5,28)(105.5,24)
 \Line(109.5,19)(112.5,15)
  \Vertex(37.5,15){2.5}
  \Vertex(37.5,90){2.5}
  \Vertex(112.5,15){2.5}
  \Vertex(112.5,90){2.5}
  \ArrowLine(150,90)(112.5,90)
  \ArrowLine(112.5,90)(112.5,15)
  \ArrowLine(112.5,15)(150,15)
\Text(18.75,105)[tc]{$e^+$}
  \Text(75,108.75)[tc]{$\phi^0$}
  \Text(131.25,112.5)[tc]{$\bar t$}
  \Text(18.75,52.5)[lc]{$e$}
  \Text(131.8,52.5)[lc]{$t$}
  \Text(18.75,0)[cb]{$e^-$}
  \Text(75,-3.75)[bc]{$A$}
  \Text(131.25,-3)[cb]{$t$}
\end{picture}
\qquad \qquad \qquad
&+&
\quad 
\begin{picture}(100,105)(0,49.5)
  \ArrowLine(0,15)(37.5,15)
  \ArrowLine(37.5,15)(37.5,90)
  \ArrowLine(37.5,90)(0,90)
  \Photon(37.5,90)(112.5,15){4}{20}
 \Line(37.5,15)(41.5,19 )
 \Line(45.5,24)(49.5,28)
 \Line(53.5,32)(57.5,36)
 \Line(61.5,40)(65.5,44)
 \Line(69.5,48)(73.5,52)
 \Line(77.5,56)(81.5,60)
 \Line(85.5,64)(89.5,68)
 \Line(93.5,72)(97.5,76)
 \Line(101.5,80)(105.5,84)
 \Line(109.5,88)(112.5,90)
  \Vertex(37.5,15){2.5}
  \Vertex(37.5,90){2.5}
  \Vertex(112.5,15){2.5} 
  \Vertex(112.5,90){2.5}
  \ArrowLine(150,90)(112.5,90)
  \ArrowLine(112.5,90)(112.5,15)
  \ArrowLine(112.5,15)(150,15)
\Text(18.75,105)[tc]{$e^+$}
  \Text(75,108.75)[tc]{$A$}
  \Text(131.25,112.5)[tc]{$\bar t$}
  \Text(18.75,52.5)[lc]{$e$}
  \Text(131.8,52.5)[lc]{$t$}
  \Text(18.75,0)[cb]{$e^-$}
  \Text(75,-3.75)[bc]{$\phi^0 $}
  \Text(131.25,-3)[cb]{$t$}
\end{picture}
\end{array}
}
\]
\vspace{10mm}
\caption{Direct and crossed $ZA$ boxes. \label{ZA_box_dc} }
\end{figure}
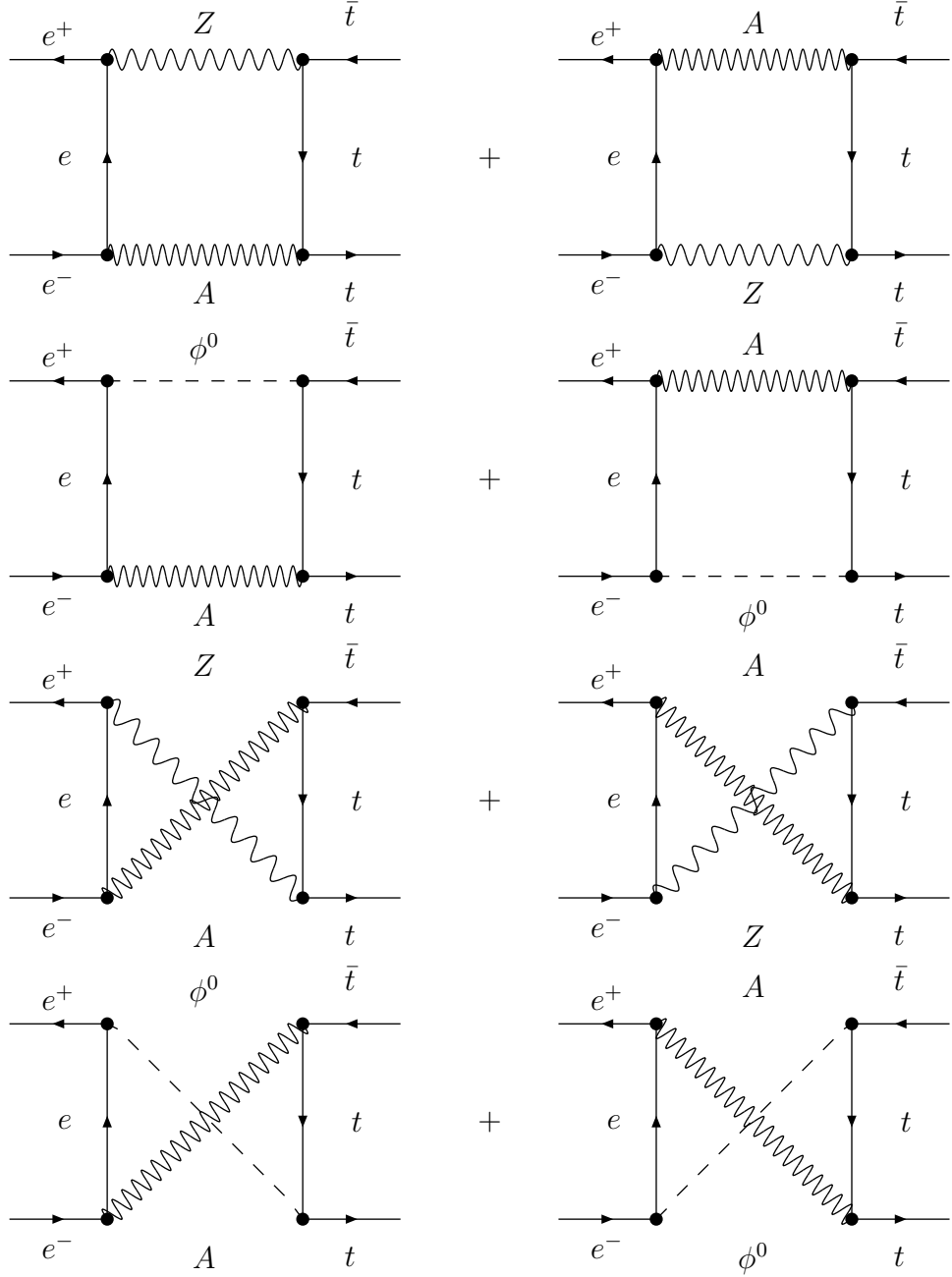

The six relevant scalar form factors are conveniently presentable in form of differences 
of $t$ and $u$ dependent functions:
\bqa
\Fza{IJ}\lpar s,t,u \rpar  &=& \Fza{IJ}(s,t)-\Fza{IJ}(s,u),
\label{FzaIJ}
\eqa
where index $IJ$ is any pair of $L,R\oplus L,R,D$. The 12 $\Fza{IJ}$ functions depend on
6 auxiliary functions by means of equations where the coupling constants are factored out:
\[
\begin{array}{rclclrclcl}
\Fza{LL}\lpar s,t \rpar &=&   \vpa{e}{} \vpa{t}{} {\cal G}_1\lpar s,t \rpar 
                &\hspace*{-2mm}+\hspace*{-2mm}& \vpa{e}{} \vma{t}{} {\cal G}_2\lpar s,t \rpar,
&
\Fza{LL}\lpar s,u \rpar &=&   \vpa{e}{} \vma{t}{} {\cal H}_1\lpar s,u \rpar 
                &\hspace*{-2mm}+\hspace*{-2mm}& \vpa{e}{} \vpa{t}{} {\cal H}_2\lpar s,u \rpar,
\nll[2mm]
\Fza{RR}\lpar s,t \rpar &=&   \vma{e}{} \vma{t}{} {\cal G}_1\lpar s,t \rpar 
                &\hspace*{-2mm}+\hspace*{-2mm}& \vma{e}{} \vpa{t}{} {\cal G}_2\lpar s,t \rpar,
&
\Fza{RR}\lpar s,u \rpar &=&   \vma{e}{} \vpa{t}{} {\cal H}_1\lpar s,u \rpar 
                &\hspace*{-2mm}+\hspace*{-2mm}& \vma{e}{} \vma{t}{} {\cal H}_2\lpar s,u \rpar,
\nll[2mm]
\Fza{LR}\lpar s,t \rpar &=&   \vpa{e}{} \vpa{t}{} {\cal H}_1\lpar s,t \rpar 
                &\hspace*{-2mm}+\hspace*{-2mm}& \vpa{e}{} \vma{t}{} {\cal H}_2\lpar s,t \rpar,
&
\Fza{LR}\lpar s,u \rpar &=&   \vpa{e}{} \vma{t}{} {\cal G}_1\lpar s,u \rpar 
                &\hspace*{-2mm}+\hspace*{-2mm}& \vpa{e}{} \vpa{t}{} {\cal G}_2\lpar s,u \rpar,
\nll[2mm]
\Fza{RL}\lpar s,t \rpar &=&   \vma{e}{} \vma{t}{} {\cal H}_1\lpar s,t \rpar 
                &\hspace*{-2mm}+\hspace*{-2mm}& \vma{e}{} \vpa{t}{} {\cal H}_2\lpar s,t \rpar,
&
\Fza{RL}\lpar s,u \rpar &=&   \vma{e}{} \vpa{t}{} {\cal G}_1\lpar s,u \rpar 
                &\hspace*{-2mm}+\hspace*{-2mm}& \vma{e}{} \vma{t}{} {\cal G}_2\lpar s,u \rpar,
\nll[2mm]
\Fza{LD}\lpar s,t \rpar &=&   \vpa{e}{} \vpa{t}{} {\cal F}_1\lpar s,t \rpar 
                &\hspace*{-2mm}+\hspace*{-2mm}& \vpa{e}{} \vma{t}{} {\cal F}_2\lpar s,t \rpar,
&
\Fza{LD}\lpar s,u \rpar &=&   \vpa{e}{} \vma{t}{} {\cal F}_1\lpar s,u \rpar 
                &\hspace*{-2mm}+\hspace*{-2mm}& \vpa{e}{} \vpa{t}{} {\cal F}_2\lpar s,u \rpar,
\nll[2mm]
\Fza{RD}\lpar s,t \rpar &=&   \vma{e}{} \vma{t}{} {\cal F}_1\lpar s,t \rpar 
                &\hspace*{-2mm}+\hspace*{-2mm}& \vma{e}{} \vpa{t}{} {\cal F}_2\lpar s,t \rpar,
&
\Fza{RD}\lpar s,u \rpar &=&   \vma{e}{} \vpa{t}{} {\cal F}_1\lpar s,u \rpar 
                &\hspace*{-2mm}+\hspace*{-2mm}& \vma{e}{} \vma{t}{} {\cal F}_2\lpar s,u \rpar.
\end{array}
\]
Finally, we present these 6 auxiliary functions:
\bqa
{\cal F}_1 \lpar s,t \rpar &\hspace*{-2mm}=\hspace*{-2mm}& -\frac{1}{8}\frac{s}{\sdfit} \bigg\{
            \tmi\bigg[\lpar \Rz+\frac{\tmi}{s}-2 \rpar \jazt
\nll[2mm] &&
               -4\cattze\bigg]
\nll[2mm] &&
               + 2\frac{\szmi\tmi}{\sdfit}
                    \bigg[ t\jazt+2 t\cattze
\nll &&
               -\tmi\ceszeo-\tpl\ctszto \bigg]
\nll &&
               - \szpl\ceszeo-\szmi\ctszto
\nll[2mm] &&
           - 2 t\, \cattze
           - 2\frac{t}{\tmi} \bigg[ \mzs \cattze
\nll &&
            -2\boftet+\boftzt+\boftto\bigg] 
\nll &&
           - 2\frac{\tpl}{\sdtit} 
         \bigg[ \lpar \mzs - 4 \mts \rpar  \ctszto
\nll &&
           + 2\bofszo-\boftzt-\boftto \bigg] \bigg\},
\\
{\cal F}_2 \lpar s,t \rpar &\hspace*{-2mm}=\hspace*{-2mm}& \frac{1}{8}\frac{s}{\sdfit} 
                  \, \bigg[\frac{ \tmi^2}{s}\jazt-2 t \cattze\bigg],
\eqa
\vspace*{-6.5mm}
\bqa
{\cal H}_1 \lpar s,t \rpar &\hspace*{-2mm}=\hspace*{-2mm}& 
      - \frac{s \mts}{8} \, \bigg\{ \frac{1}{s} \bigg[\jazt-2\frac{\mzs}{\tmi}\cattze\bigg]
\nll &&
            + \frac{1}{\sdfit} \bigg(-t \bigg[ \jazt + 2 \cattze\bigg] 
\nll &&
            + 4\mts \ctszto
            + \tmi \lpar \Rz-2+\frac{\szmi\tpl}{\sdfit}\rpar
                                    \bigg[\jazt
\nll &&
     +2\cattze - 2 \ctszto\bigg]
\nll &&
            + \szmi \lpar s+2\tmi \rpar \frac{\tmi}{\sdfit} 
               \bigg[ \ctszto 
\nll &&
-\ceszeo \bigg]   
            - \szpl\ceszeo
\nll [1.5mm] &&
            + \lpar \Rz-1\rpar \lpar s+2\tmi \rpar\ctszto
\nll [2.5mm] &&
            - 2\bofszo+2\boftet
\\[1mm] &&
            + 2\frac{\mts}{\tmi} \bigg[2\boftet-\boftzt
                                       -\boftto \bigg] \bigg) \bigg\},
\nll[2mm]
{\cal H}_2 \lpar s,t \rpar &\hspace*{-2mm}=\hspace*{-2mm}& 
s \, \bigg\{ -\frac{\tmi}{4 \szmi} \bigg[\jazt-\cateot
\nll &&
+\cattze\bigg]
        +\frac{\mts}{8} 
 \bigg(\frac{1}{s}\jazt
\nll &&
-\frac{\mts}{\sdfit} \bigg[\jazt+2\cattze\bigg] \bigg) \bigg\},
\eqa
\vspace*{-3mm}
\bqa
{\cal G}_1 \lpar s,t \rpar &\hspace*{-2mm}=\hspace*{-2mm}& 
s \, \bigg\{ \frac{\tmi}{4 \szmi} \bigg[-\jazt+\cateot
\nll &&
-\cattze\bigg]
            -\frac{\mts}{8 s} \jazt 
\nll &&
            + \frac{1}{8 \sdfit} \bigg(
            + \mtq \bigg[ \jazt + 2\cattze\bigg]
\nll &&
            - \tmi \bigg[ -\bigg( 2\tpl-\frac{\mts\mzs}{s}\bigg) \jazt
\nll && 
            - 4\tpl\cattze + 2\mts\ctszto \bigg]
\nll &&
            - \frac{\szmi\tmi\tpl}{\sdfit}
                \bigg[ t\jazt+2t\cattze
\nll &&
            - \tpl\ctszto -\tmi\ceszeo \bigg]
\nll &&
            + t \bigg[\szpl\ceszeo
            +\lpar\szpl-4\mts\rpar\ctszto
\nll &&
                   +2\bofszo-2\boftet \bigg] 
\nll &&
               + 2\frac{\mts t}{\tmi}\bigg[\mzs\cattze
\nll &&
           - 2 \boftet
           +   \boftzt+\boftto\bigg] \bigg) \bigg\},
\\
{\cal G}_2 \lpar s,t \rpar &\hspace*{-2mm}=\hspace*{-2mm}& \frac{ s \mts}{8} \, 
\bigg[ \frac{1}{s}\jazt
\nll &&
-\frac{t}{\sdfit}
       \bigg( \jazt+2\cattze \bigg) \bigg],
\eqa
where new notation were introduced for invariants
\bq
s_{\pm}= s \pm \mzs\,,
\qquad
t_{\pm}= t \pm \mts\,,
\eq
and for the new functions $J^{\sss IJ}\lpar Q^2,P^2;M_1,M_2 \rpar$
(an analog of $J^{\sss AA}\lpar Q^2,P^2;M_1,M_2 \rpar$ \eqn{JAA}):
\bqa
J^{\sss AZ}\lpar Q^2,P^2;M_1,M_2 \rpar
&=& \frac{1}{P^2+M^2_2} 
    \ln\bigg(\frac{ Q^2 + \mzs}{\mzs}\bigg)
    \ln\bigg[\frac{M^2_1 M^2_2}{\lpar P^2+M^2_2 \rpar^2} \bigg].
\eqa 
\subsubsection{Box--Born interferences}
Any box, describing by the amplitude~\eqn{anyboxamplitude}, interfering with $\gamma$ and $\zb$ 
exchange tree level amplitudes, gives rise to two contributions to the differential cross-sections,
which are useful for internal cross-checks:
\bqa
  \sigma_{{\sss{\rm BOX}}\otimes{\sss {\rm BORN_\gamma}}} & \propto & 8 \qe \qt 
 {\rm Re} \biggl\{
    \bigg(
         \bigg[ \lpar s+\tmi \rpar^2+s\mts \bigg] 
                  \lpar {\cal F}_{\sss LL}+{\cal F}_{\sss RR} \rpar
\nll &&
       + \lpar s\mts+\tmi^2 \rpar 
                  \lpar {\cal F}_{\sss LR}+{\cal F}_{\sss RL} \rpar
       -2\mts \lpar s t+\tmi^2 \rpar 
          \lpar {\cal F}_{\sss LD}+{\cal F}_{\sss RD} \rpar \bigg) \biggr\},
\label{BoxBorngamma}
\\
  \sigma_{{\sss{\rm BOX}}\otimes{{\sss {\rm BORN}}_{\sss Z}}} & \propto &   
\label{BoxBornZ}
  8 {\rm Re} \biggl\{\bigg(
  \bigg[\lpar s+\tmi \rpar^2+s\mts\bigg] \vma{t}{}
    \lpar \vpae {\cal F}_{\sss LL}+\vma{e}{} {\cal F}_{\sss RR} \rpar
\\ &&
       +2 \lpar s+\tmi \rpar^2 \au            \vpae
                                  {\cal F}_{\sss LL}
       +2\tmi^2\au                \vma{e}{} {\cal F}_{\sss RL}
       + \lpar s \mts+\tmi^2 \rpar \vma{t}{}     
                 \lpar\vpae {\cal F}_{\sss LR}+\vma{e}{}{\cal F}_{\sss RL}\rpar
\nll &&
       +2 s \mts \au 
            \lpar \vpae {\cal F}_{\sss LR}+\vma{e}{} {\cal F}_{\sss RR}\rpar
       -2 \mts \lpar s t+\tmi^2 \rpar \vu     
              \lpar \vpae {\cal F}_{\sss LD}+\vma{e}{}{\cal F}_{\sss RD}\rpar           
                            \bigg) {\chi_{\sss Z}}^* \biggr\}.
\nonumber
\eqa
\section{Total scalar form factors of the one-loop amplitude}
\eqnzero

 Adding all contributions together, we observe the cancellation of all poles.
The ultraviolet-finite results for six scalar form factors, replacing EW result 
{\bf Eq.~(I.3.118)}, are:
\bqa
\vvertil{}{\sss{LL}}{s,t,u}&=&
    \lrbr \cvetril{zee}{\sss{L}}{s} + \cvetril{A,e}{}{s} \rrbr
 +  \cvetril{ztt}{\sss{L}}{s}  
 +  \cvertil{ct}{\sss{LL}}{s}  
 +  16k\cvetril{\sss BOX}{\sss LL}{s,t,u},
\nll [2mm]
\vvertil{}{\sss{QL}}{s,t,u}&=&
    \lrbr \cvetril{zee}{\sss{Q}}{s} + \cvetril{A,e}{}{s} \rrbr
 +  \cvetril{ztt}{\sss{L}}{s}             
 + k~\cvetril{\gamma tt}{\sss{L}}{s}
 +  \cvertil{ct}{\sss{QL}}{s}
 +  16k\cvetril{\sss BOX}{\sss QL}{s,t,u},
\nll [2mm]
\vvertil{}{\sss{LQ}}{s,t,u}&=&
     \lrbr \cvetril{zee}{\sss{L}}{s} + \cvetril{A,e}{}{s} \rrbr
            + \cvetril{ztt}{\sss{Q}}{s}    
+  k~\cvetril{\gamma ee}{\sss{L}}{s} + \cvertil{ct}{\sss{LQ}}{s}
 +  16k \cvetril{\sss BOX}{\sss LQ}{s,t,u},
\nll [1mm]
\vvertil{}{\sss{QQ}}{s,t,u}&=&
    \lrbr \cvetril{zee}{\sss{Q}}{s} + \cvetril{A,e}{}{s} \rrbr
 +  \cvetril{ztt}{\sss{Q}}{s}                   
\nll [1mm]
&&- \frac{k}{\stws}~\lrbr \cvetril{\gamma ee}{\sss{Q}}{s} + \cvetril{A,e}{}{s} 
 +  \cvetril{\gamma tt}{\sss{Q}}{s} \rrbr
 +  \cvertil{ct}{\sss{QQ}}{s}
 +  16k\cvetril{\sss BOX}{\sss QQ}{s,t,u},
\nll [1mm]
\vvertil{}{\sss{LD}}{s,t,u}&=&
    \cvetril{ztt}{\sss{D}}{s}
 +  16k\cvetril{\sss BOX}{\sss LD}{s,t,u},
\nll [4mm]
\vvertil{}{\sss{QD}}{s,t,u}&=&
    \cvetril{ztt}{\sss{D}}{s}
+ k~\cvetril{\gamma tt}{\sss{D}}{s}
 +  16k\cvetril{\sss BOX}{\sss QD}{s,t,u},
\label{soular_ff}
\eqa
where
\bq
k = \ctws \lpar \Rz-1 \rpar.
\eq
For $IJ=LL$ component of box contribution one has: 
\bq
 \cvetril{\sss BOX}{\sss {IJ}}{s,t,u}  = 
  k^{\sss AA} \cvetril{\sss AA}{\sss{IJ}}{s,t,u}
+ k^{\sss ZA} \cvetril{\sss ZA}{\sss{IJ}}{s,t,u} 
+ k^{\sss ZZ} \cvetril{\sss ZZ}{\sss{IJ}}{s,t,u}
+ k^{\sss WW} \cvetril{\sss WW}{\sss{IJ}}{s,t,u}
\label{com_box}
\eq
and for the other components $IJ=LQ,QL,QQ,LD,QD$ of box form factors 
the $WW$ box does not contribute. Moreover,
\bqa
\cvetril{\gamma(z)tt}{\sss{L,Q,D}}{s}=
\sum_{\sss B=A,Z,H,W}
\cvetril{\gamma(z)B}{\sss{L,Q,D}}{s}\,,    
\eqa
except $\cvetril{\gamma A}{\sss{L}}{s} =0$ and $\cvetril{\gamma H}{\sss{L}}{s} =0$.

\section{Process $eett$ in the helicity amplitudes}
\eqnzero

  According to the analysis of the EW part in~\cite{eett_subm} and presentation of the QED 
part here, we have the {\bf complete} answer for the amplitude of our process.
  
  The aim of this section is to adapt the helicity amplitude technigues for
the description of our process. 
  We produced an alternative analityc answer for the same amplitude
using the method suggested by Vega and Wudka (VW) ~\cite{hel_VegaWudka}.

  In general, there are 16 helicity amplitude for any $2 \ff \to 2 \ff $ process. For the unpolarized case
and when the electron mass is ignored, we are left with six independent helicity amplitudes,
which depend on kinematical variables and our six form factors: 
\bqa
{\cal A}_{++++}  &=& 0,\qquad
{\cal A}_{+++-}   =  0,\qquad
{\cal A}_{++-+}   =  0,\qquad 
{\cal A}_{++--}   =  0,
\nll
{\cal A}_{+-+-}  &=& s\lpar 1-\cos\vartheta\rpar 
 \bigg( \qe \qt F_{\sss GG}
        +{\chi_{\sss Z}}\delta_e \bigg[\lpar 1+\beta_t\rpar \tcit F_{\sss QL}
                              +\delta_t F_{\sss QQ}\bigg]\bigg),
\nll
{\cal A}_{+--+}  
                  &=& s \lpar 1+\cos\vartheta\rpar 
 \bigg(  \qe\qt F_{\sss GG}
        +{\chi_{\sss Z}}\delta_e\bigg[(1-\beta_t)\tcit  F_{\sss QL}
        +\delta_t F_{\sss QQ}\bigg] \bigg),
\nll
{\cal A}_{+---} &=& {\cal A}_{+-++}
                   =  2\sqs\mtl \sin\vartheta
 \bigg(\qe \qt F_{\sss GG}
       +{\chi_{\sss Z}}\delta_e \bigg[ \tcit F_{\sss QL}+\delta_t  F_{\sss QQ}
     +\frac{1}{2} s {\beta_t}^2  \tcit  F_{\sss QD}\bigg]\bigg),
\nll
{\cal A}_{-+++} &=& {\cal A}_{-+--} 
                   =  -2\sqs\mtl \sin\vartheta
 \bigg(\qe\qt F_{\sss GG}
\nll &&
 +{\chi_{\sss Z}} \bigg[2 \tcie \tcit F_{\sss LL}+2 \tcie  \delta_t F_{\sss LQ}
 +\delta_e\tcit F_{\sss QL}
                                 +\delta_e \delta_t F_{\sss QQ}
\nll &&
  +\frac{1}{2} s {\beta_t}^2 \tcit
       \lpar 2\tcie F_{\sss LD}+\delta_e F_{\sss QD}\rpar\bigg]\bigg),
\nll
{\cal A}_{-++-} &=&  s \lpar 1+\cos\vartheta \rpar 
  \bigg( \qe\qt F_{\sss GG}
\nll &&
         +{\chi_{\sss Z}}\bigg[ \lpar 1+\beta_t \rpar 
    \lpar 2\tcie \tcit F_{\sss LL}+\delta_e \tcit  F_{\sss QL} \rpar
+\delta_t\lpar 2 \tcie F_{\sss LQ}+\delta_e  F_{\sss QQ} \rpar \bigg]\bigg),
\nll
{\cal A}_{-+-+} &=&  s \lpar 1-\cos\vartheta \rpar 
   \bigg(
  \qe\qt F_{\sss GG}
\nll &&
+{\chi_{\sss Z}} \bigg[\lpar 1-\beta_t\rpar \tcit
        \lpar 2\tcie    F_{\sss LL}+\delta_e F_{\sss QL}\rpar
             +\delta_t \lpar 2 \tcie F_{\sss LQ}
             +\delta_e  F_{\sss QQ} \rpar \bigg] \bigg),
\nll
{\cal A}_{--++} &=& 0,\qquad
{\cal A}_{--+-}  =  0,\qquad
{\cal A}_{---+}  =  0,\qquad
{\cal A}_{----}  =  0.
\eqa
Here
\bqa
\cos\vartheta &=& \lpar t-\mts+\frac{s}{2} \rpar \frac{2}{s\beta_t},
\eqa
and  for the amplitude 
$ {\cal A}_{ \lambda_{i} \lambda_{j} \lambda_{k} \lambda_{l} }  $
each index $ \lambda_{(i,j,k,l)}$ 
takes two values $\pm$ meaning twice projection of spins 
$e^+, e^-, t, \bar{t}$ onto their corresponding momentum.
The differential cross-section for the unpolarized case is:
\bqa
\frac{d\sigma}{d\cos\vartheta} = \,\frac{\pi \alpha^2}{s^3}\beta_t N_c \,
 \sum_{ \lambda_{i} \lambda_{j} \lambda_{k} \lambda_{l}}
\left|
{\cal A}_{ \lambda_{i} \lambda_{j} \lambda_{k} \lambda_{l} }  
\right|^2.
\label{dsigma_A}
\eqa

We checked, that this expression is analytically identical to {\bf Eq.~(I.4.122)}.
 The expression \eqn{dsigma_A} contains, however, spurious
contributions of the two-loop order (squares of one-loop terms), which one should 
supress, since we would like to have 
a complete one-loop result. 

 This may be achieved with a simple trick.
 First of all let us note, that if all form factors are: $ F_{\sss IJ} = 1 $ 
for $IJ = LL,\,LQ,\,QL,\,QQ$ and $F_{\sss IJ} = 0$ for $IJ = LD,\,QD$, we have the tree level.
 At the one-loop level $LL,\,LQ,\,QL,\,QQ$ form factors may be represented as:
\bqa 
{\bf F}_{\sss IJ} 
= 1 + \frac{\alpha}{4 \pi \siw^2} F_{\sss IJ}\,, 
\label{one_ll}
\eqa
and
\bqa 
{\bf F}_{\sss IJ} 
= \frac{\alpha}{4 \pi \siw^2} F_{\sss IJ}\,, 
\label{one_llD}
\eqa
for $IJ = LD,\,QD$.

Instead of \eqn{one_ll} for the four form factors we write 
\bqa 
{\bf F}_{\sss IJ} 
= Z + \frac{\alpha}{4 \pi \siw^2} F_{\sss IJ}\,,
\label{Pl_z}
\eqa
and note that the cross section is a function of six form factors.
 
 Then the one-loop results apparently equals:
\bqa
\frac{{d\sigma}^{(1)}}{d\cos\vartheta} = 
 \frac{d\sigma}{d\cos\vartheta}[Z=1] - \frac{d\sigma}{d\cos\vartheta}[Z=0]. 
\label{oloop}
\eqa

\section{QED annex}
\eqnzero

\subsection{QED vertices and soft photon contributions}
Here we present virtual corrections due to QED vertices, a factorised part due to QED boxes
and soft photon contributions. 
The expressions of this subsection can be also casted from \cite{Bardin:1999ak}.

The formal structure of factorised virtual and soft contributions is as follows:
\bqa
\delta^{\sss{\rm virt+soft}} = 
\label{dlt_vs} 
\frac{\alpha}{\pi} 
 \bigg[ \qe^2 \delta^{\sss{\rm virt+soft}}_{\sss{\rm ISR}} 
      + \qe\qt\delta^{\sss{\rm virt+soft}}_{\sss{\rm IFI}} 
      + \qt^2 \delta^{\sss{\rm virt+soft}}_{\sss{\rm FSR}} \bigg].
\eqa
There are three types of contributions: ISR, FSR and IFI.
\subsubsection{Initial state radiation (ISR)}
Contributions of the initial state QED $e^+e^-\gamma$ vertex and ISR soft are short, since 
electron mass is ignored:
\bqa
\delta^{\rm virt}_{\sss{\rm ISR}} &=& 
            - \ln \frac{\mes}{\tHlas} 
                    \bigg( \lelog   - 1 \bigg)
            - \frac{1}{2}  \lelog^2 
            + \frac{3}{2}  \lelog   - 2 + 4 \Litwo\lpar 1 \rpar,
\nll
\delta^{\rm soft}_{\sss{\rm ISR}}    &=& 
\ln\lpar \frac{4 \omega^2}{s} \frac{\mes}{\tHlas} \rpar 
             \bigg(  \lelog - 1 \bigg)
       + \frac{1}{2} \lelog^2  -  2 \Litwo\lpar 1 \rpar,
\label{isr_virt}
\eqa
where
\bq
l_e  = \ln \lpar \frac{s}{\mes} \rpar.
\eq
\subsubsection{Initial--final state interference (IFI)} 
This originates from contributions of QED boxes: \,$\gamma\gamma, \zb\gamma$ and 
initial--final state soft photons interference:
\bqa      
\delta^{\rm virt}_{\sss{\rm IFI}} 
\label{dltIFI}
&=& - 2 \ln \frac{s}{\tHlas} \ln \frac{\tmi}{\umi},
\\
\delta^{\rm soft}_{\sss{\rm IFI}} &=& 2 \ln \frac{4\omega^2}{\tHlas} \ln\frac{\tmi}{\umi}
\label{dltsft}
                  +\lrbr F^{\rm soft} \lpar s,t \rpar - F^{\rm soft}\lpar s,u \rpar \rrbr,
\eqa  
with
\bqa
F^{\rm soft} \lpar s,t \rpar &=& 
         - \frac{1}{2} \lelog^2
         - \frac{1}{2} \ln^2\eta
         +2  \ln\eta
             \ln  \lpar 1+\frac{2 \mts}{\betatp\tmi}\rpar
\\&&
         -   \ln^2\lpar 1+\frac{2 \mts}{\betatp\tmi}\rpar
         +   \ln^2\lpar-\frac{s t}{\tmis}\rpar
         +2  \ln\lpar-\frac{s t}{\tmis}\rpar \ln \lpar 1+\frac{\tmis}{s t}\rpar
\nll &&
         +2  \Litwo\lpar 1 - \frac{2 t}{\tmi\betatp} \rpar
         -2  \Litwo\lpar\frac{-\betatm\tmi}{\betatp\tmi+2\mts} \rpar
         -2  \Litwo\lpar-\frac{\tmis}{ s t}\rpar
         -2  \Litwo\lpar 1 \rpar,
\nonumber
\eqa
where we introduce the notations
\bq
\beta \equiv \beta_t = \sqrt{1-\frac{4 \mts}{s}}\;,
\qquad\betatp = 1+\beta\,,\qquad\betatm = 1-\beta\,,
\qquad\eta    = \frac{\betatm}{\betatp}\;.
\eq
\subsubsection{Final state radiation (FSR)}
Contributions of one-loop QED $ f \bar f \gamma$ vertex and 
final state soft photon radiation are: 
\bqa
\delta^{\sss{\rm virt}}_{\rm{\sss FSR}}
        &=&-\ln\frac{\mts}{\tHlas} \bigg[-\bebeta \ln\eta-1\bigg]
                           -\frac{3}{2}\beta\ln\eta-2
\nll &&
       +\bebeta \bigg[-\frac{1}{2} \ln^2\eta
                + 2 \ln\eta \ln\lpar 1-\eta\rpar
                + 2 \Litwo \lpar\eta\rpar + 4 \Litwo\lpar 1 \rpar \bigg],
\nll
\delta^{\sss{\rm soft}}_{\rm{\sss FSR}}
   &=& \ln\frac{4\omega^2}{\tHlas}\bigg[-\bebeta\ln\eta-1 \bigg]
        -\frac{1}{\beta} \ln\eta
\label{penka}
\nll &&
       +\bebeta \bigg[-\frac{1}{2} \ln^2\eta
                + 2 \ln\eta \ln\lpar 1-\eta\rpar
               + 2 \Litwo\lpar\eta\rpar-2 \Litwo\lpar 1 \rpar\bigg].
\eqa
 Contribution of the ISR ~\eqn{isr_virt} may be received from these expressions
in the limit $ m_t=m_e \to 0$.

\subsubsection{Non-factorized final state vertex `anomalous' contributions}

 For presentation of this contribution let us introduce the definition
\bqa
 L_n=\ln\frac{\beta-1}{\beta+1}.
\label{Leta}
\eqa
The `anomalous' part of QED vertex contribution to the differential cross-section
reads:
\bqa
\frac{d\sigma^a}{d\cos\vartheta} 
&=&
  4 \alpha^3 N_c \frac{\mts}{s^4} \qts \Bigg[
          \Bigg( \qes \qts + 2 \qe\qt v_e v_t Re({\chi_{\sss Z}})
       + \lpar v^2_e + a^2_e \rpar v^2_t  |{\chi_{\sss Z}}|^2 \Bigg)
           \lpar s t+\tmi^2 \rpar Re(L_n)
\nll &&
       + \qe \qt a_e a_t s \lpar s + 2 \tmi \rpar Re(L_n {\chi_{\sss Z}})
\nll &&
       + \Bigg(
         \lpar v^2_e + a^2_e  \rpar  a^2_t
        \lrbr  s \lpar s-4\mts \rpar+  2\lpar  s t+\tmi^2 \rpar \rrbr
\nll &&
 + 2 v_e a_e v_t a_t s \lpar s + 2 \tmi\rpar  \Bigg) |{\chi_{\sss Z}}|^2 Re(L_n) \Bigg].
\eqa

\subsection{An alternative form of the cross-section for QED boxes}
Here we present some useful formulae which are not in the main stream of our approach
(described in previous Sections), but that were used for internal cross checks of calculations
of the QED part of the process under consideration.

The QED boxes \eqnsc{BoxBorngamma}{BoxBornZ} may be greatly simplified purely algebraically.
For the sum of $\ab\ab$ and $\zb\ab$ boxes one may easily derive the cross-section:
\bqa
\frac{d\sigma^{\sss{\rm BOX}} }{d\cos\vartheta} 
&=& \frac{2\alpha^3}{s}\beta \qe\qt N_c \, {\rm Re} \,
  \biggl\{
                \qe^2 \qt^2 \, {\cal F}_{\sss V}                           
  +\qe \qt \, {\chi_{\sss Z}} \bigg[v_e v_t \left({\cal F}^*_{\sss V} +{\cal H}_{\sss V} \right) 
  +                      a_e a_t \left({\cal F}^*_{\sss A} +{\cal G}_{\sss V} \right) \bigg]  
\nll && 
  +|{\chi_{\sss Z}}|^2    \bigg[
\left( v_e^2 + a_e^2 \right) \lpar \,v_t^2 {\cal H}_{\sss V} + a_t^2  {\cal H}_{\sss A} \rpar  
  +2 a_e v_e a_t v_t \, \lpar {\cal G}_{\sss V} + {\cal G}_{\sss A} \rpar 
                          \bigg] \biggr\},
\label{alernativeQED}
\eqa    
where $\chi_{\sss{Z}}(\sman)$ is defined by \eqn{propagators} and the six {\em cross-section
form fartors} are:
\bqa
{\cal F}_{\sss V}  &=& {\cal F}_{\sss V} \lpar t \rpar - {\cal F}_{\sss V} \lpar u \rpar,
\nll 
{\cal F}_{\sss A}  &=& {\cal F}_{\sss A} \lpar t \rpar + {\cal F}_{\sss A} \lpar u \rpar,
\nll 
{\cal H}_{\sss V}  &=& {\cal H}_{\sss V} \lpar t \rpar - {\cal H}_{\sss V} \lpar u \rpar,
\nll 
{\cal H}_{\sss A}  &=& {\cal H}_{\sss A} \lpar t \rpar - {\cal H}_{\sss A} \lpar u \rpar,
\nll 
{\cal G}_{\sss V}  &=& {\cal G}_{\sss V} \lpar t \rpar + {\cal G}_{\sss V} \lpar u \rpar,
\nll 
{\cal G}_{\sss A}  &=& {\cal G}_{\sss A} \lpar t \rpar + {\cal G}_{\sss A} \lpar u \rpar,
\eqa
with
\bqa
{\cal F}_{\sss V} \lpar t \rpar &=& \frac{1}{s} \biggl\{
        \frac{\tmi}{4} \bigg[ 2 \mts+\lpar s + 2\tmi \rpar\bigg] \jaat
\nll &&
      + t \bigg( \frac{1}{2}  
          \bigg[ s \cesoeo + \lpar s-4\mts \rpar \ctsoto \bigg]
\nll &&
       +\frac{2 \mts}{\sdtit} \bigg[ 2 \mts \ctsoto+\boftto-\bofsoo \bigg] \bigg)             
\nll &&
      - \frac{ s \mts}{\tmi}  \bigg[ \boftet-\boftto \bigg]
\nll &&
      - \frac{\lpar s+\tmi \rpar }{2} \bigg[\boftet-\bofsoo\bigg] 
    \bigg\},
\\[2mm]
{\cal F}_{\sss A} \lpar t \rpar &=& \frac{1}{s} \bigg\{ 
        \frac{ s + 2\tmi}{4} 
 \tmi \jaat - \mts  \bigg(\frac{1}{2} s \ctsoto
\nll &&
           +\lpar \frac{s}{\tmi}+1 \rpar \bigg[ \boftet-\boftto \bigg]
                    \bigg)
\nll &&
      -\frac{\lpar s+\tmi \rpar}{2} \bigg[ \boftet-\bofsoo \bigg] \bigg\},
\\
{\cal H}_0 \lpar t \rpar &=& \frac{1}{s^2}\bigg\{ 
     -  \lpar \tmi^2+\lpar s+\tmi \rpar^2 \rpar 
            \tmi    \bigg[ \jazt - \frac{1}{2}\jaat
\nll &&
                 + \cattze \bigg] 
\nll &&
       + \szmi \bigg(
  \tmi \lpar \szpl + 2 t \rpar \bigg[\frac{1}{2}\jazt + \cattze \bigg]       
\nll &&
       + \frac{s \mts}{\tmi}\bigg[ \mzs \cattze - 2 \boftet
\nll &&
       + \boftto + \boftzt  \bigg]
\nll &&
       + s t \bigg[  \ceszeo + \ctszto \bigg]
\nll &&
       - \lpar s+\tmi \rpar \bigg[\boftet - \bofszo \bigg] \bigg) \bigg\},
\\[2mm]
{\cal H}_{\sss V} \lpar t \rpar &=& {\cal H}_0 \lpar t \rpar + 
     \frac{2 \mts}{s^2}\bigg\{ - s \tmi   \bigg[ \jazt - \frac{1}{2}\jaat
\nll &&
       + \cattze \bigg]
\nll &&
       + \szmi t \bigg( -\ctszto        
       + \frac{1}{\sdtit} \bigg[  
         \szmi \ctszto
\nll &&
      +  \boftto+\boftzt-2 \bofszo \bigg] \bigg) \bigg\},
\\[2mm]
{\cal H}_{\sss A} \lpar t \rpar &=& {\cal H}_0 \lpar t \rpar +
   \frac{2 \mts}{s^2}            \bigg\{
   \tmi \bigg[\mzs\jazt
\nll &&
         - s \lpar \frac{1}{2}\jaat - \cattze\rpar\bigg]     
\nll &&
         + \szmi \bigg[-t \ctszto + \szpl\cattze
\nll &&
         - \boftet+\bofszo\bigg] \bigg\}, 
\\[2mm]
{\cal G}_{\sss V} \lpar t \rpar &=&
   - \frac{1}{s} \bigg\{
         \lpar s + 2\tmi \rpar \tmi \bigg[ \jazt 
\nll &&
        - \frac{1}{2}\jaat + \cattze \bigg]
\nll &&
        -  \szmi \bigg( \frac{\tmi}{s} \lpar \szpl +  2\tmi \rpar 
                          \bigg[ \frac{1}{2}\jazt+\cattze \bigg]
\nll &&
        + \frac{1}{2} \mzs \bigg[ \ceszeo+\ctszto \bigg]
\nll &&
        + \mts \bigg( 2  \cattze
        + \frac{1}{\tmi} \bigg[\mzs \cattze
\nll &&
      -2 \boftet  + \boftto+\boftzt\bigg]
\nll &&
        - \frac{\lpar s+\tmi \rpar}{s} \bigg[\boftet-\bofszo\bigg]
   \bigg) \bigg\},
\\[2mm]
{\cal G}_{\sss A} \lpar t \rpar &=& {\cal G}_{\sss V} \lpar t \rpar -
        \mts \frac{\szmi}{s^2}    
  \bigg[ 2 \szmi \cattze 
\nll &&
+ \szpl \ctszto 
\nll &&
+ 2 \boftet - \boftzt - \boftto \bigg].
\label{lffn}
\eqa  
The \eqns{alernativeQED}{lffn} were coded as a separate branch of {\tt eettLib} and together
with vertex QED contributions described in the previous subsections was used for 
internal cross-check of QED part of calculations.

 Some factorized part of the $AA$ and $ZA$ boxes contribution is not included in ~\eqn{alernativeQED}.
It has a form
\bqa
\frac{d\sigma}{d\cos\vartheta} \frac{\alpha}{\pi}\qe\qt \delta^{\rm virt}_{\sss{\rm IFI}},
\eqa
where $\delta^{\rm virt}_{\sss{\rm IFI}} $ is given by ~\eqn{dltIFI}.

 The whole QED contribution can be written as follows
\bqa
\frac{d\sigma^{\sss {\rm QED}}}{d\cos\vartheta}=
\frac{d\sigma^{\sss{\rm BORN}}}{d\cos\vartheta}\delta^{\rm virt+soft}
+\frac{d\sigma^a}{d\cos\vartheta}
+\frac{d\sigma^{\sss{\rm BOX}} }{d\cos\vartheta}, 
\eqa
where $\delta^{\rm virt+soft} $ is defined by \eqns{dlt_vs}{penka}. 
\newpage
\section{Numerical results and discussion}
\eqnzero

All the formulae derived in this paper as well as in Ref.~\cite{eett_subm} are realized in a 
{\tt FORTRAN} code with a tentative name {\tt eeffLib}. Numbers presented in this section are 
produced with updated, February 2002, version of the code. As compared to December 2000 version,
used to produce numbers for Ref.~\cite{eett_subm}, current version contains full QED corrections 
together with the soft photon contribution to the angular distribution $d\sigma/d\cos\vartheta$. 
Morever, two bugs of December 2000 version were fixed, which resulted in a change of numbers. 
First, for light final state fermion masses, the numerical precision was being lost. 
After curing this oddity, the agreement between {\tt eeffLib} and {\tt ZFITTER} numbers became 
even better. Secondly, there was a bug in a part of {\tt FORTRAN} code computing $\zb\zb$ box 
contribution. Its fixing resulted in a change of numbers (in 4th--5th digits) for the case of 
heavy final state fermion masses (top quark). 
Since numbers which were presented in Ref.~\cite{eett_subm} are changed anyway, we decided 
to present again all the Tables that were already given in Ref.~\cite{eett_subm}. 
On top of it, we will show several new examples of numbers. In particular, we will show
a comparison of the electroweak form factors (EWFF) {\em including} QED corrections between 
{\tt eeffLib} and another {\tt FORTRAN} code, which was automatically generated 
from {\tt form} log files with the aid of 
a system {\tt s2n\_f} ({\it symbolic to numbers}), producing a {\tt FORTRAN} source code 
--- a part of our {\tt CalcPHEP} system. 
This comparison provides a powerful internal 
cross-check of our numerics that practically excludes appearance of bugs of a kind discussed above.

We begin with showing several examples of comparison with 
{\tt ZFITTER v6.30}~\cite{zfitterv6.30:2000}. In the present realization,
{\tt eeffLib} does not calculate $\mwl$ from $\mu$ decay and does not precompute either
Sirlin's parameter $\Delta r$ or total $\zb$ width, which enters the $\zb$ boson propagator.
For this reason, the three parameters: $\mwl\,,\;\Delta r\,,\;\gz$ were being taken from
{\tt ZFITTER} and used as {\tt INPUT} for {\tt eeffLib}. Moreover, present {\tt eeffLib}
is a purely one-loop code, while in {\tt ZFITTER} it was not foreseen to access just one-loop
form factors with users flags. To accomplish the goals of comparison at the one-loop level, 
we had to modify the {\tt DIZET} electroweak library. The most important change 
was an addition to the {\tt SUBROUTINE ROKANC}:
\vspace*{-5mm}

\begin{verbatim}
*
* For eett
*  
      FLL=(XROK(1)-1D0+DR )*R1/AL4PI
      FQL=FLL+(XROK(2)-1D0)*R1/AL4PI
      FLQ=FLL+(XROK(3)-1D0)*R1/AL4PI
      FQQ=FLL+(XROK(4)-1D0)*R1/AL4PI 
\end{verbatim}
with the aid of which we reconstruct four form factors from {\tt ZFITTER}'s effective
couplings $\rho$ and $\kappa$'s ($F_{\sss{LD}}$ and $F_{\sss{QD}}$ do not contribute 
in massless approximation).
\subsection{Flags of {\tt eeffLib}\label{t-flaggs}}
Here we give a description of flags (user options) of {\tt eeffLib}. While creating 
the code, we followed the  principle to preserve as much as possible the meaning of flags as 
described in the {\tt ZFITTER} description~\cite{Bardin:1999yd}. In the list below, 
a comment `{\tt as in ZFD}' means that the flag has exactly the same meaning as in
~\cite{Bardin:1999yd}. Here we describe an extended set of flags of February 2002 version
of {\tt eeffLib}.
\begin{itemize}
\item \verb+ALEM=3  ! as in ZFD +
\vspace*{-2.5mm}

\item \verb+ALE2=3  ! as in ZFD +
\vspace*{-2.5mm}

\item \verb+VPOL=0  ! =0 + $\alpha$(0); =1,=2 as in ZFD; =3 is reserved for later use \\
Note that the flag is extended to {\tt VPOL=0} to allow calculations `without running of 
$\alpha$'.
\vspace*{-2.5mm}

\item \verb+QCDC=0  ! as in ZFD +
\vspace*{-2.5mm}

\item \verb+ITOP=1  ! as in DIZET (internal flag) +
\vspace*{-2.5mm}

\item \verb+GAMS=1  ! as in ZFD +
\vspace*{-2.5mm}

\item \verb+WEAK=1  ! as in ZFD + (use {\tt WEAK=2} in v6.30 to throw away some higher order terms)
\vspace*{-2.5mm}

\item \verb+IMOMS=1 ! =0 +  $\alpha$-scheme; =1 GFermi-scheme  \\
New meaning of an old flag: switches between two renormalization schemes;
\vspace*{-2.5mm}

\item \verb+BOXD=6 + \\
Together with {\tt WEAK=0} is used for an internal comparison 
of separate boxes and QED contributions: \\
      \verb+BOXD    ! =1 + with $\ph\ph$ boxes \\
      \verb+        ! =2 + with $\zb\ph$ boxes \\
      \verb+        ! =3 + with $\ph\ph$ and $\zb\ph$ boxes \\ 
      \verb+        ! =4 + with all QED contributions       \\
Together with {\tt WEAK=1} (working option), it has somewhat different meaning: \\
      \verb+BOXD    ! =0 + without any boxes   \\
      \verb+        ! =1 + with $\ph\ph$ boxes \\
      \verb+        ! =2 + with $\zb\ph$ boxes \\
      \verb+        ! =3 + with $\ph\ph$ and $\zb\ph$ boxes \\ 
      \verb+        ! =4 + with $\wb\wb$ boxes \\
      \verb+        ! =5 + with $\wb\wb$ and $\zb\zb$ boxes \\
      \verb+        ! =6 + with all QED and EW boxes 
\vspace*{+2mm}

{\bf `Treatment' options.}

\item \verb+GAMZTR=1+ treatment of $\Gamma_{\sss{Z}}$. \\
The option is implemented for the sake of comparison with 
{\tt FeynArts}: \\
      \verb+GAMZTR=0 + $\Gamma_{\sss{Z}} =  0 $ \\
      \verb+GAMZTR=1 + $\Gamma_{\sss{Z}}\ne 0 $
\vspace*{-2.5mm}

\item \verb+EWFFTR=0+ treatment of EW form factors. \\
Switches between form factors and 
effective {\tt ZFITTER} couplings $\rho$ and $\kappa$'s. The option is implemented for
comparison with {\tt ZFITTER}: \\
      \verb+EWFFTR=0 + electroweak form factors \\
      \verb+EWFFTR=1 + effective coullings $\rho$ and $\kappa$
\vspace*{-2.5mm}

\item \verb+FERMTR=1+ treatment of fermionic masses.\\
Switches between three different
sets of `effective quark masses':\\
      \verb+FERMTR=1 + a `standard' set of fermions masses\\
      \verb+FERMTR=2,3+  `modified' sets 
\vspace*{-2.5mm}

\item \verb+VPOLTR=1+ treatment of photonic vacuum polarization.\\ 
Switches between
lowest order expression, $\alpha(s)=\alpha\lrbr 1+\Delta\alpha(s)\rrbr$, and its 
`resummed' version, $\alpha(s)=\alpha/\lrbr 1-\Delta\alpha(s)\rrbr$:\\
      \verb+VPOLTR=0 + lowest order\\
      \verb+VPOLTR=1 + resummed
\vspace*{-2.5mm}
 
\item \verb+EWRCTR=2+ treatment of electroweak radiative corrections.\\ 
Switches between three 
variants for vertex corrections:\\
      \verb+EWRCTR=0 + electroweak form factors contain only QED additions  \\
      \verb+EWRCTR=1 + electroweak form factors do not contain QED additions\\ 
      \verb+EWRCTR=2 + electroweak form factors contain both QED and EW additions
\vspace*{-2.5mm}

\item \verb+EMASTR=0+ treatment of terms with $\ln(s/\mes)$ in $\ph\ph$ and $\zb\ph$ boxes,
which are present in various functions but cancel in sum:\\
      \verb+EMASTR=0 + these terms are suppressed in all functions which they enter \\
      \verb+EMASTR=1 + these terms are retained in all functions which results in loosing
of computer precision owing to numerical cancellation; results for \verb+EMASTR=0+ and 
\verb+EMASTR=1+ are equal 
\vspace*{-7mm}

\item \verb+EWWFFV=1+ treatment of vertex and box diagrams with virtual $\wb$ boson,
switches between two variants:\\
      \verb+EWWFFV=0 + variant of formulae without $b$-quark mass\\
      \verb+EWWFFV=1 + variant of formulae with finite $b$-quark mass
\vspace*{+2mm}

{\bf Options affecting QED contributions.}

\item \verb+IQED=4 + variants of inclusion virtual and soft photon QED contributions:\\
      \verb+IQED=1 + only initial state radiation (ISR)     \\
      \verb+IQED=2 + only initial--final interference (IFI) \\
      \verb+IQED=3 + only ~final~ state radiation (FSR)     \\
      \verb+IQED=4 + all QED contributions are included     
\vspace*{-2.5mm}

\item \verb+IBOX=4 + is active only if {\tt IQED}=2 or 4 and affects only Eq.~(5.11): \\
      \verb+IBOX=0 + AA boxes interfering with $\ph$ exchange BORN \\
      \verb+IBOX=1 + AA boxes \\
      \verb+IBOX=2 + ZA boxes \\
      \verb+IBOX=3 or 4 + AA+ZA boxes 
\end{itemize}

\subsection{{\tt eeffLib}--{\tt ZFITTER} comparison of scalar form factors}
First of all we discuss the results of a computation of the four scalar form factors, 
\bqa
\vvertil{}{\sss{LL}}{s,t},\quad
\vvertil{}{\sss{QL}}{s,t},\quad
\vvertil{}{\sss{LQ}}{s,t},\quad
\vvertil{}{\sss{QQ}}{s,t},
\eqa
for three variants:\\
1) without EW boxes, i.e. without gauge-invariant contribution of $\zb\zb$ boxes, 
and without 
$\gpar=1$ part of the $\wb\wb$ box, Eq.~{\bf (I.2.93)};\\
2) with $\wb\wb$ boxes;\\
3) with $\wb\wb$ and $\zb\zb$ boxes.

\begin{table}[!t]
\vspace*{-1cm}
\caption[EWFF for the process $\fep\fem\to\fu\bar{u}$. {\tt eeffLib}--{\tt ZFITTER} 
comparison without and with $\wb\wb$ boxes.]
{EWFF for the process $\fep\fem\to\fu\bar{u}$. {\tt eeffLib}--{\tt ZFITTER} comparison.
\label{t_com1}}
\vspace*{3mm}
\begin{tabular} {||c|c||l|l|l||}
\hline
\hline
\multicolumn{5}{||c||}{Without EW boxes}                                               \\
\hline
\multicolumn{2}{||c||}{$\sqrt s$}&~~~~~~~~100 GeV    &~~~~~~~~200 GeV    &~~~~~~~~300 GeV \\
\hline
FF&$\mu$& & & \\
\hline
\hline
           &$\wml/10$&$13.47773~-~i1.84784$&$16.22034~-i10.49412$&$23.75240~-i11.27469$\\
$F_{\sss{LL}}$&$\wml$&$13.47773~-~i1.84784$&$16.22034~-i10.49412$&$23.75240~-i11.27469$\\
           &$10\wml $&$13.47773~-~i1.84784$&$16.22034~-i10.49412$&$23.75240~-i11.27469$\\
\hline
\multicolumn{2}{||c||}{\tt ZFITTER}
                     &$13.47771~-~i1.84786$&$16.22031~-i10.49405$&$23.75237~-i11.27464$\\
\hline
\hline
           &$\wml/10$&$29.34721~+~i3.67330$&$30.33891~+~i3.34531$&$31.64553~+~i2.75258$\\
$F_{\sss{QL}}$&$\wml$&$29.34721~+~i3.67330$&$30.33891~+~i3.34531$&$31.64553~+~i2.75258$\\
           &$10\wml $&$29.34721~+~i3.67330$&$30.33891~+~i3.34531$&$31.64553~+~i2.75258$\\
\hline
\multicolumn{2}{||c||}{\tt ZFITTER}
                     &$29.34720~+~i3.67330$&$30.33889~+~i3.34535$&$31.64552~+~i2.75259$\\
\hline
\hline
           &$\wml/10$&$29.13302~+~i3.26972$&$30.03854~+~i1.54158$&$31.68636~-~i0.22635$\\
$F_{\sss{LQ}}$&$\wml$&$29.13302~+~i3.26972$&$30.03854~+~i1.54158$&$31.68636~-~i0.22635$\\
           &$10\wml $&$29.13302~+~i3.26972$&$30.03854~+~i1.54158$&$31.68636~-~i0.22635$\\
\hline
\multicolumn{2}{||c||}{\tt ZFITTER}
                     &$29.13304~+~i3.26973$&$30.03855~+~i1.54163$&$31.68635~-~i0.22634$\\
\hline
\hline
           &$\wml/10$&$44.90390~+~i8.85688$&$43.80286~+i10.02412$&$44.21223~+i10.83899$\\
$F_{\sss{QQ}}$&$\wml$&$44.90390~+~i8.85688$&$43.80286~+i10.02412$&$44.21223~+i10.83899$\\
           &$10\wml $&$44.90390~+~i8.85688$&$43.80286~+i10.02412$&$44.21223~+i10.83899$\\
\hline
\multicolumn{2}{||c||}{\tt ZFITTER}
                     &$44.90392~+~i8.85688$&$43.80285~+i10.02411$&$44.21224~+i10.83894$\\
\hline
\hline
\multicolumn{5}{||c||}{$\wb\wb$ is added}                                             \\
\hline
\hline
           &$\wml/10$&$12.94469~-~i1.84784$&$~9.34066~-~i9.42482$&$~9.03908~-i11.55971$\\
$F_{\sss{LL}}$&$\wml$&$12.94469~-~i1.84784$&$~9.34066~-~i9.42482$&$~9.03908~-i11.55971$\\
           &$10\wml $&$12.94469~-~i1.84784$&$~9.34066~-~i9.42482$&$~9.03908~-i11.55971$\\
\hline
\multicolumn{2}{||c||}{\tt ZFITTER}
                     &$12.94468~-~i1.84786$&$~9.34065~-~i9.42467$&$~9.03903~-i11.55958$\\
\hline
\hline
\end{tabular}
\end{table}

In this comparison we use flags as in subsection \ref{t-flaggs} and, moreover,
\bqa
\mwl    &=&80.4514958\;\mbox{GeV},
\nl
\Delta r&=&0.0284190602\,,
\nl
\Gamma_{\sss Z}&=&2.499 776\;\mbox{GeV}.
\label{firstPOs}
\eqa
In \tbn{t_com1} we show an example of comparison of four form factors 
$\vvertil{}{\sss{LL,QL,LQ,QQ}}{\sman,\tman}$ between the {\tt eeffLib}, where we set 
$\mtl=0.2$
 GeV and {\tt ZFITTER} (the latter is able to deliver only massless results).
The form factors are shown as complex numbers for the three c.m.s. energies 
(for $\tman = \mts-\sman/2$)
and for the three values of scale $\tHs=\wml/10,\;\wml,\;10\wml$. The table demonstrates 
scale independence and very good agreement with {\tt ZFITTER} results (6 or 7 digits).
One should stress that total agreement with {\tt ZFITTER} is not expected because 
in the {\tt eeffLib} code we use massive expressions to compute the nearly massless case.
Certain numerical cancellations leading to losing some numerical precision are expected.  
We should conclude that the agreement is very good and uniquely demonstrates that our 
formulae have the correct $\mtl\to 0$ limit.
\newpage

\begin{table}[!t]
\caption[EWFF for the process $\fep\fem\to\fu\bar{u}$. {\tt eeffLib}--{\tt ZFITTER} comparison
with $\zb\zb$ boxes.]
{EWFF for the process $\fep\fem\to\fu\bar{u}$. {\tt eeffLib}--{\tt ZFITTER} comparison.
\label{t_com2}}
\vspace*{2mm}
\begin{tabular} {||c|c||l|l|l||}
\hline
\hline
\multicolumn{5}{||c||}{With $\zb\zb$ boxes}                                            \\
\hline
\multicolumn{2}{||c||}{$\sqrt s$}&~~~~~~~~100 GeV    &~~~~~~~~200 GeV   &~~~~~~~~300 GeV \\
\hline
FF&$\mu$& & & \\
\hline
\hline
           &$\wml/10$&$12.89584~-~i1.84784$&$~8.24736~-i10.64666$&$~8.98375~-i12.88478$\\
$F_{\sss{LL}}$&$\wml$&$12.89584~-~i1.84784$&$~8.24736~-i10.64666$&$~8.98375~-i12.88478$\\
           &$10\wml $&$12.89584~-~i1.84784$&$~8.24736~-i10.64666$&$~8.98375~-i12.88478$\\
\hline
\multicolumn{2}{||c||}{\tt ZFITTER}
                     &$12.89583~-~i1.84786$&$~8.24736~-i10.64651$&$~8.98370~-i12.88466$\\
\hline
\hline
           &$\wml/10$&$29.30447~+~i3.67330$&$29.38219~+~i2.27610$&$31.59711~+~i1.59302$\\
$F_{\sss{QL}}$&$\wml$&$29.30447~+~i3.67330$&$29.38219~+~i2.27610$&$31.59711~+~i1.59302$\\
           &$10\wml $&$29.30447~+~i3.67330$&$29.38219~+~i2.27610$&$31.59711~+~i1.59302$\\
\hline
\multicolumn{2}{||c||}{\tt ZFITTER}
                     &$29.30445~+~i3.67330$&$29.38216~+~i2.27613$&$31.59710~+~i1.59304$\\
\hline
\hline
           &$\wml/10$&$29.10829~+~i3.26972$&$29.48510~+~i0.92307$&$31.65836~-~i0.89713$\\
$F_{\sss{LQ}}$&$\wml$&$29.10829~+~i3.26972$&$29.48510~+~i0.92307$&$31.65836~-~i0.89713$\\
           &$10\wml $&$29.10829~+~i3.26972$&$29.48510~+~i0.92307$&$31.65836~-~i0.89713$\\
\hline
\multicolumn{2}{||c||}{\tt ZFITTER}
                     &$29.10832~+~i3.26973$&$29.48512~+~i0.92312$&$31.65835~-~i0.89711$\\
\hline
\hline
           &$\wml/10$&$44.88226~+~i8.85688$&$43.31855~+~i9.48287$&$44.18773~+i10.25200$\\
$F_{\sss{QQ}}$&$\wml$&$44.88226~+~i8.85688$&$43.31855~+~i9.48287$&$44.18773~+i10.25200$\\
           &$10\wml $&$44.88226~+~i8.85688$&$43.31855~+~i9.48287$&$44.18773~+i10.25200$\\
\hline
\multicolumn{2}{||c||}{\tt ZFITTER}
                     &$44.88228~+~i8.85688$&$43.31854~+~i9.48286$&$44.18773~+i10.25196$\\
\hline
\hline
\end{tabular}
\end{table}
\vspace*{-8mm}

In \tbn{t_com2} we show a similar comparison with {\tt ZFITTER} when $\zb\zb$ boxes are added.
As seen, the agreement has not deteriorated.
\subsection{{\tt eeffLib}--{\tt ZFITTER} comparison of IBA cross-section}
As the next step of the comparison of {\tt eeffLib} with calculations from the 
literature, we present a comparison of the IBA cross-section.

In \tbn{IBA_table1} we show the differential cross-section Eq.~(4.44) in pb
for three values of $\cos\vartheta=-0.9,\,0,\,+0.9$, with input parameters
of \eqn{firstPOs} and with constant e.m. coupling $\alpha=\alpha(0)$.

Next, we present the same comparison as in \tbn{IBA_table1}, but now with running e.m. 
coupling.
Since the flags setting {\tt VPOL=1}, which is  relevant to this case, affects {\tt ZFITTER}
numbers, we now use, instead of \eqn{firstPOs}, the new {\tt INPUT} set:
\bqa
\mwl    &=&80.4467671\;\mbox{GeV},
\nl
\Delta r&=&0.0284495385\,,
\nl
\Gamma_{\sss Z}&=&2.499 538\;\mbox{GeV}.
\label{secondPOs}
\eqa

\begin{table}[t]
\caption[{\tt eeffLib}--{\tt ZFITTER} comparison of the differential cross-section
without running $\alpha$.]
{IBA, First  row -- {\tt ZFITTER} ($u\bar{u}$ channel); second 
row -- {\tt eeffLib} ($\mtl=0.1$ GeV, $\mbl=0$ GeV); 
 third  row -- {\tt eeffLib} ($\mtl=173.8$ GeV); with constant e.m. coupling $\alpha=\alpha(0)$.}
\label{IBA_table1}
\vspace*{5mm}
\centering
\begin{tabular} {||c|l|l|l|l|l|l||}
\hline
\hline
${\sqrt s}$         & 100  GeV & 200 GeV  & 300 GeV  & 400 GeV  & 700 GeV  & 1000 GeV\\
\hline
                    & 47.664652& 0.291823 & 0.169510 &          &          &          \\ 
\cline{2-7}
$\cos\vartheta=-0.9$& 47.661843& 0.291827 & 0.169515 & 0.103284 & 0.035319 & 0.017204 \\
\cline{2-7}
                    &          &          &          & 0.162193 & 0.044088 & 0.018927 \\
\hline
\hline
                    & 59.768387& 1.718830 & 0.695061 &          &          &          \\
\cline{2-7}
$\cos\vartheta= 0 $ & 59.770299& 1.718870 & 0.695075 & 0.376871 & 0.117279 & 0.055873 \\
\cline{2-7}
                    &          &          &          & 0.264713 & 0.112918 & 0.054209 \\
\hline
\hline
                    &168.981978& 5.954048 & 2.292260 &          &          &          \\
\cline{2-7}
$\cos\vartheta= 0.9$&168.991144& 5.954182 & 2.292302 & 1.222354 & 0.372912 & 0.176038 \\
\cline{2-7}
                    &          &          &          & 0.438453 & 0.293399 & 0.154785 \\
\hline
\hline
\end{tabular}
\end{table}

\begin{table}[h]
\caption[{\tt eeffLib}--{\tt ZFITTER} comparison of the differential cross-section
with running $\alpha$.]
{IBA, First  row -- {\tt ZFITTER} ($u\bar{u}$ channel); second row -- {\tt eeffLib} 
($\mtl=0.1$ GeV, $\mbl=0$ GeV); third  row -- {\tt eeffLib} ($\mtl=173.8$ GeV); 
with running e.m. coupling $\alpha=\alpha(s)$.}
\label{IBA_table2}
\vspace*{5mm}
\centering
\begin{tabular} {||c|l|l|l|l|l|l||}
\hline
\hline
${\sqrt s}$       &  100 GeV  & 200 GeV  & 300 GeV  & 400 GeV  & 700 GeV  & 1000 GeV \\
\hline
                  &  45.404742& 0.386966 & 0.225923 &          &          &          \\
$\cos\vartheta=-0.9$&45.404598& 0.386966 & 0.225923 & 0.138065 & 0.048621 & 0.024156 \\
                  &           &          &          & 0.194752 & 0.058013 & 0.025959 \\
\hline
\hline
                  &  60.382423& 1.882835 & 0.771939 &          &          &          \\
$\cos\vartheta= 0 $& 60.382562& 1.882837 & 0.771939 & 0.421410 & 0.133475 & 0.064245 \\
                  &           &          &          & 0.303683 & 0.130173 & 0.062838 \\
\hline
\hline
                  &  173.467517&6.450000 & 2.510881 &          &          &          \\
$\cos\vartheta= 0.9$&173.467543&6.450000 & 2.510881 & 1.346620 & 0.417295 & 0.198842 \\
                  &            &         &          & 0.492546 & 0.330401 & 0.175564 \\

\hline
\hline
\end{tabular}
\end{table}

The numbers, shown in first two rows of \tbnsc{IBA_table1}{IBA_table2} 
exhibit a very good level of agreement for light quark masses, while the third rows illustrate
the mass effect due to heavy top.

Finally, in \tbn{IBA_table3}, we give a comparison of the cross-section integrated within  
the angular interval $|\cos\vartheta| \leq 0.999$. (Flags setting is the same as for Table 4.)

A typical deviation between {\tt eeffLib} and {\tt ZFITTER} is of the order $\sim 10^{-6}$, i.e. 
of the order of the required precision of the numerical integration over $\cos\vartheta$.
Examples of numbers obtained with {\tt eeffLib}, which were shown in this section, 
demonstrate that {\tt ZFITTER} numbers are recovered for light $\mtl$.

\begin{table}[t]
\caption[{\tt eeffLib}--{\tt ZFITTER} comparison of the total cross-section.]
{{\tt eeffLib(L)}--{\tt ZFITTER(Z)} comparison of the total cross-section.
Cross-sections are given in pb:
the first row  -- $\sigma^{L}_{\rm{tot}}$, i.e. {\tt eeffLib} ($m_t=0.1$ GeV); 
the second row -- $\sigma^{Z}_{\rm{tot}}$, i.e. {\tt ZFITTER} ($u\bar{u}$ channel); 
the third row shows the absolute deviation $\displaystyle{\sigma^{L}_{\rm{tot}}
                                                         -\sigma^{Z}_{\rm{tot}}}$.}
\label{IBA_table3}
\vspace*{5mm}
\centering
\begin{tabular}{||c|c||c|c||c|c||}
\hline
\hline
\multicolumn{2}{||c||}{$100$ GeV}&\multicolumn{2}{|c||}{200 GeV}  
                                 &\multicolumn{2}{|c||}{300 GeV}       \\
\hline
$\sigma_{\rm{tot}}$&$\sigma_{\sss{\rm{FB}}}$&$\sigma_{\rm{tot}}$&$\sigma_{\sss{\rm{FB}}}$
                                  &$\sigma_{\rm{tot}}$&$\sigma_{\sss{\rm{FB}}}$\\
\hline
  160.8981   &  70.98416   &   5.021810   &   3.360848   &    2.031754   &    1.269556  \\
  160.8980   &  70.98406   &   5.021808   &   3.360848   &    2.031754   &    1.269556  \\
\hline 
 ~~~0.0001   &  ~0.00010   &   0.000002   &   0.0        &    0.0        &    0.0       \\
\hline
\hline
\end{tabular}
\end{table}
We conclude this subsection with a comment about technical precision of our calculations.
We do not use {\tt looptools} package~\cite{Hahn:2000jm}.
For all PV functions, but one, namely $D_0$ function, we use our own coding where we can
control precision internally and, typically, we can guarantee 11 digits precision.
For $D_0$ function we use, instead, {\tt REAL*16 TOPAZ0} coding~\cite{Montagna:1999kp} and 
the only accessible for us way to control the precision is to compare results with 
those computed with  {\tt looptools} package.
This was done for a typical $D_0$ functions entering $ZZ$ box contributions.
We got an agreement within 14-15 digits between these two versions for all 
$\sqrt{s}=400 - 10000$ GeV and $\cos\vartheta=0.99,\,0,\,-0.99$.

\subsection{Comparison with a code generated by {\tt s2n\_f}}
Here we present a numerical comparison of the complete scalar form factors \eqn{soular_ff} 
extracted from two independently created codes: `manually written' {\tt eettLib}
and a code, `automatically generated' by {\tt s2n\_f} software.
We use a special input parameter set here: all lepton masses $\alpha$ and a conversion factor
from GeV$^{-1}$ to pb are taken from 2000 of Particle Data Tables while for quark and
photon and gauge boson masses we use:
\bqa
m_{u,d,c,s,t,b} &=& 0.062,\;0.083,\;1.50,\;0.215,\;173.8,\;4.70\;\mbox{GeV},
\nl
\lambda&=&1\;\mbox{GeV},\;\mzl=zm=91.1867\;\mbox{GeV},\;\mwl=80.4514958\;\mbox{GeV}.
\eqa
\begin{table}[!h]
\vspace*{-5mm}
\tiny
\caption[EWFF for the process $\fep\fem\to\ft\bar{t}$. {\tt eettLib}--{\tt s2n\_f} comparison.]
{EWFF for the process $\fep\fem\to\ft\bar{t}$. 
{\tt eettLib}, first rows; {\tt s2n\_f}, second rows.
\label{t_com6}}
\vspace*{3mm}
\begin{tabular}{||c|c|c|c|c||}
\hline
\hline
\multicolumn{2}{||c|}{$\sqrt{s}$}& 400 GeV & 700 GeV & 1000 GeV       \\
\hline
$\cos\vartheta$&FF&  &  &        \\
\hline
\hline
-0.9&$F_{\sss{LL}}$
 &$68.36399900074-i1.24743850729   $&$79.63957322115-i20.53758995637  $&$80.47816819240-i26.71016937527  $\\
&&$68.36399900068-i1.24743850728   $&$79.63957322113-i20.53758995637  $&$80.47816819239-i26.71016937527  $\\
    &$F_{\sss{QL}}$
 &$75.12465846647+i34.81991916400  $&$76.19283172015+i28.44336684106  $&$75.95332822621+i27.77201429453  $\\
&&$75.12465846641+i34.81991916400  $&$76.19283172013+i28.44336684106  $&$75.95332822620+i27.77201429453  $\\ 
    &$F_{\sss{LQ}}$
 &$81.01546270426+i19.81343626967  $&$82.67283873006+i13.79952080171  $&$83.26485989744+i12.23741074712  $\\
&&$81.01546270420+i19.81343626968  $&$82.67283873004+i13.79952080171  $&$83.26485989743+i12.23741074712  $\\
    &$F_{\sss{QQ}}$
 &$225.63977621858+i154.37838168488$&$207.09189805263+i133.45188150116$&$194.07155316803+i134.33226297675$\\
&&$225.63977621832+i154.37838168491$&$207.09189805254+i133.45188150117$&$194.07155316799+i134.33226297675$\\
    &$F_{\sss{LD}}$
 &$ -0.57522852857+i0.34010611241~~$&$ -0.33030593699+i0.14897150833~~$&$-0.22418674728-i0.08847119487~~$\\
&&$ -0.57522852857+i0.34010611241~~$&$ -0.33030593699+i0.14897150833~~$&$-0.22418674728-i0.08847119487~~$\\
    &$F_{\sss{QD}}$
 &$ 0.16677424366-i0.34326069364   $&$  0.29925308488-i0.14107543098  $&$ 0.23436559470-i0.05839137636  $\\
&&$ 0.16677424366-i0.34326069364   $&$  0.29925308488-i0.14107543098  $&$ 0.23436559470-i0.05839137636  $\\
\hline
 0.0&$F_{\sss{LL}}$
 &$48.42950001713+i8.26103890366   $&$ 28.23570422021+i2.43705570966  $&$ 16.54896558498+i0.77082434583 $\\
&&$48.42950001707+i8.26103890367   $&$ 28.23570422019+i2.43705570966  $&$ 16.54896558497+i0.77082434583 $\\
&$F_{\sss{QL}}$
 &$68.02678564355+i37.08805801477  $&$ 58.00469565609+i33.82433896562 $&$ 52.56343218854+i34.28418004972$\\
&&$68.02678564349+i37.08805801477  $&$ 58.00469565607+i33.82433896562 $&$ 52.56343218853+i34.28418004972$\\
&$F_{\sss{LQ}}$
 &$73.37133716227+i22.69397728402  $&$ 62.40775508619+i20.75544388763 $&$ 56.94788960099+i20.65389886886$\\
&&$73.37133716220+i22.69397728403  $&$ 62.40775508616+i20.75544388764 $&$ 56.94788960098+i20.65389886886$\\
&$F_{\sss{QQ}}$
 &$196.60425612149+i162.74818773960$&$132.63279537966+i152.68259938740$&$~98.94491326876+i157.45863002555$\\
&&$196.60425612123+i162.74818773963$&$132.63279537957+i152.68259938741$&$~98.94491326872+i157.45863002556$\\
&$F_{\sss{LD}}$
 &$ -0.56319765502+i0.33645326768~~$&$ -0.29067043403+i0.13992893252~~$&$ -0.18096486789+i0.08187546112~~$\\
&&$ -0.56319765502+i0.33645326768~~$&$ -0.29067043403+i0.13992893252~~$&$ -0.18096486789+i0.08187546112~~$\\
&$F_{\sss{QD}}$
 &$ 0.15893936555-i0.37254018572   $&$  0.26429138671-i0.15437851127  $&$  0.18981891199-i0.06181088399  $\\
&&$ 0.15893936555-i0.37254018572   $&$  0.26429138671-i0.15437851127  $&$  0.18981891199-i0.06181088399  $\\
\hline
 0.9&$F_{\sss{LL}}$
 &$35.17736865724+i14.84038724783\,$&$\;0.21531292996+i13.66645015866 $&$-18.22896674792+i13.10520552155~\,$\\
&&$35.17736865718+i14.84038724784\,$&$\;0.21531292994+i13.66645015866 $&$-18.22896674793+i13.10520552155~\,$\\
    &$F_{\sss{QL}}$
 &$61.03099608330+i39.09196533610  $&$ 40.77942026097+i37.94118444135 $&$  30.73687048410+i39.06896169999$\\
&&$61.03099608324+i39.09196533611  $&$ 40.77942026095+i37.94118444136 $&$  30.73687048409+i39.06896169999$\\
    &$F_{\sss{LQ}}$
 &$66.08215572935+i25.04151178684  $&$ 44.50915974057+i25.51875704261 $&$  34.09235431695+i26.17012224125$\\
&&$66.08215572929+i25.04151178685  $&$ 44.50915974055+i25.51875704261 $&$  34.09235431694+i26.17012224125$\\
    &$F_{\sss{QQ}}$
 &$167.63393504156+i170.36384103672$&$~59.87568281297+i168.13599380718$&$ ~~7.10290370391+i175.24101109592$\\
&&$167.63393504130+i170.36384103675$&$~59.87568281288+i168.13599380719$&$ ~~7.10290370387+i175.24101109593$\\
    &$F_{\sss{LD}}$
 &$ -0.56772633347+i0.34299744419~~$&$ -0.32035310873+i0.14419510235~~$&$  -0.21547870582+i0.08254457292~~$\\
&&$ -0.56772633347+i0.34299744419~~$&$ -0.32035310873+i0.14419510235~~$&$  -0.21547870582+i0.08254457292~~$\\
    &$F_{\sss{QD}}$
 &$ 0.18031346246-i0.40091423652   $&$  0.34968026058-i0.16945266925  $&$   0.29109057806-i0.06775284666  $\\
&&$ 0.18031346246-i0.40091423652   $&$  0.34968026058-i0.16945266925  $&$   0.29109057806-i0.06775284666  $\\
\hline
\hline
\end{tabular}
\vspace*{-7mm}
\end{table}
\clearpage

As seen from the table numbers agree within 11--13 digits, i.e. {\tt REAL*8} computational
precision is saturated.
Form factors $F_{\sss{LD,QD}}$ are multiplied  by $10^4$ to make more digits visible.

Next Table demonstrates {\tt eettLib}--{\tt s2n\_f} comparison for the complete 
one-loop differential cross-sections ${d\sigma^{(1)}}/{d\cos\vartheta}$ for the standard input set.
As seen, numbers agree within 12--13 digits.

\begin{table}[h]
\vspace*{-5mm}
\caption[$\frac{\ds d\sigma^{(1)}}{\ds d\cos\vartheta}$ for the process 
$\fep\fem\to\ft\bar{t}$. {\tt eettLib}--{\tt s2n\_f} comparison.]
        {$\frac{\ds d\sigma^{(1)}}{\ds d\cos\vartheta}$ for the process 
$\fep\fem\to\ft\bar{t}$. {\tt eettLib}--{\tt s2n\_f} comparison.}
\label{t2_p2}
\vspace*{3mm}
\centering
\begin{tabular}{||c|c|c|c||}
\hline
\hline
 $\sqrt{s}$   &        400.0        &         700.0        &      1000.0        
\\
\hline
$\cos\vartheta$ &  &  &        \\
\hline
-0.900   &   0.22357662754774  &    0.06610825350063  &  0.02926006442715   \\
$\qquad$ &   0.22357662754769  &    0.06610825350063  &  0.02926006442715
\\                         
\hline 		   					    
 0.000   &   0.34494634728716  &    0.14342802645636  &  0.06752160108814   \\
$\qquad$ &   0.34494634728707  &    0.14342802645634  &  0.06752160108813
\\
\hline 			   					    
 0.900   &   0.54806778978208  &    0.33837133344667  &  0.16973989931024   \\
$\qquad$ &   0.54806778978194  &    0.33837133344664  &  0.16973989931023
\\
\hline 
\hline 
\end{tabular} 
\vspace*{-5mm}
\end{table}

\subsection{About a comparison with the other codes}
As is well known, the one-loop differential cross-section of $\fep\fem\to\ft\bar{t}$
may be generated with the aid of the FeynArts system~\cite{Hahn:2000jm}.
Previous attempt to compare with FeynArts are described in~\cite{eett_subm}.
In December 2001, we were provided with the numbers computed with the FeynArts 
system~\cite{FynArts:2000} for $d\sigma/d\cos\vartheta$ with and without QED contributions
at $\sqrt{\sman}=700$ GeV and three values of $\cos\vartheta=0.9,\;,0,\;-0.9$.
After debugging of our code {\tt eettLib}, as described in the beginning of this section,
we eventually reached 11 digits agreement both for the tree level and 
one-loop corrected cross-sections.

We do not update Fig.13 and Fig.14 of~\cite{eett_subm}, since 
the differences with updated version is not seen.

\begin{table}[h]
\vspace*{-5mm}
\caption[$\frac{\ds d\sigma^{(1)}}{\ds d\cos\vartheta}$ for the process 
$\fep\fem\to\ft\bar{t}$ with soft photons, $E^{\rm max}_\gamma = \sqrt{s}/10$.]
{ $\frac{\ds d\sigma^{(1)}}{\ds d\cos\vartheta}$ for the process 
$\fep\fem\to\ft\bar{t}$ with soft photons, $E^{\rm max}_\gamma = \sqrt{s}/10$.}
\label{t8}
\vspace*{3mm}
\centering
\begin{tabular}{||c|c|c|c||}
\hline
\hline
 $\sqrt{s}$   &        400.0        &         700.0        &      1000.0        
\\
\hline
$\cos\vartheta$ &  &  &        \\
\hline
-0.900 &  0.17613018248935  &    0.05199100267864 &  0.02310170508071 \\
\hline
-0.500 &  0.21014509428358  &   0.06560630503586  &  0.02882301902010 \\
\hline
 0.000 &  0.27268108572063  &   0.11496514450150  &  0.05495088904853 \\
\hline
 0.500 &  0.35592722356682  &   0.19615154401629  &  0.09941700898317 \\
\hline
 0.900 &  0.43637377538440  &   0.27915043976042  &  0.14426233253975 \\
\hline 
\hline 
\end{tabular} 
\end{table}

Recently, a Bielefeld--Zeuthen team~\cite{Zeuthen:2001}
performed an alternative calculations using the DIANA system~\cite{DIANA}.
Working in close contact with this team, we managed to perform several high-precision
comparisons reaching for separate contributions an agreement in 10 digits.

The results of a comparison between FeynArts and Bielefeld--Zeuthen team are 
presented in detail in \cite{K-BZcomparison:2002}.

 As another example we present in \tbn{t8} the same cross-section
$\big[{ d\sigma}/{ d\cos\vartheta}\big]_{\sss {\rm SM}}$ as given in tables 
of~\cite{K-BZcomparison:2002}.
 For the complete cross-section, including soft photons, we agree with
Bielefeld--Zeuthen calculations within 7-8 digits. 

\addcontentsline{toc}{section}{Acknowledgments}
\section*{Acknowledgements}
 We would like to thank W.~Hollik and C.~Schappacher for a discussion of issues of the 
comparison with FeynArts.
 We acknowledge a common work on numerical comparison with J.~Fleischer, A.~Leike, T.~Riemann, 
and A.~Werthenbach which helped us to debug our `manually written' code {\tt eettLib}.
 We also wish to thank G.~Altarelli for extending to us the hospitality of the CERN TH Division 
at various stages of this work.

%% file: eett_refs_P2.tex
\addcontentsline{toc}{section}{References}
$\;$

\vspace*{-3cm}
{

}